\documentclass[twocolumn]{revtex4}
\usepackage[colorlinks,bookmarksopen,bookmarksnumbered,citecolor=magenta,urlcolor=red]{hyperref}
\usepackage{graphicx}
\usepackage{amsmath}
\usepackage{natbib}
\usepackage{subfigure}
\usepackage{xcolor}
\usepackage[percent]{overpic}
\usepackage{tikz}
\usepackage{amsmath}
\usepackage{url}
\usepackage{times}
\usepackage{epstopdf}
\begin{document}
\bibliographystyle{abbrvnat}
	
\title{Surface Mixing by Geostrophic Flows in the Bay of Bengal}
	
	
	

\author{Nihar Paul and Jai Sukhatme}

\affiliation{Centre for Atmospheric and Oceanic Sciences, Indian Institute of Science,\\
	Bangalore 560012, India}	
	
	
\begin{abstract}
		
Mixing of passive tracers in the Bay of Bengal, driven by altimetry derived daily geostrophic surface currents,
is studied on subseasonal timescales.
To begin with, Hovm{\"o}ller plots, wavenumber-frequency diagrams and power spectra confirm the multiscale nature of the flow.
Advection of latitudinal and longitudinal bands immediately brings out the chaotic nature of mixing in the Bay via
repeated straining and filamentation of the tracer field. A principal finding is that mixing is local, i.e., of the scale of the eddies,
and does not span the entire basin. Indeed, Finite Time Lyapunov Exponent (FTLE), Relative Dispersion (RD) and Finite Size Lyapunov Exponents (FSLE) maps in all seasons are patchy with minima scattered through the interior of the Bay. Further, FTLE, FSLE and RD maps show that the Bay experiences a seasonal cycle wherein rapid stirring progressively moves from the northern to southern Bay during pre and post monsoonal periods, respectively. The non-uniform stirring of the Bay is reflected in long tailed histograms of FTLEs, that become more stretched for longer time intervals. Quantitatively, advection for a week shows the mean FTLE lies near 0.15-0.16 $day^{-1}$, while extremes reach almost 0.5 $day^{-1}$. Averaged over the Bay, RD initially grows exponentially, this is followed by a power-law at scales between approximately 100 and 250 $km$, which finally transitions to an eddy-diffusive regime. These findings are confirmed by FSLEs; in addition, quantitatively, below 250 $km$, a scale dependent diffusion coefficient is extracted that behaves as a power-law with cluster size, while above 250 $km$, 		eddy-diffusivities range from $6 \times 10^3$ - $10^4$ $m^2/s$. Finally, in concert with satellite salinity data, these Lagrangian tools are used to analyse a single post-monsoonal fresh water mixing event. Here, while stirring the salinity field at large scales, FTLEs and FSLEs allow the identification of transport barriers, and elucidate how individual eddies help preserve the identity of fresh water.

\end{abstract}
	
\vskip 0.25 truecm

\maketitle
\section{Introduction}
		
Advective transport and mixing is an important aspect of geophysical flows \citep{weiss}. For example, in the oceans, surface stirring plays a key role in
determining the fate of chemical and biological fields. Stirring affects dispersal from localized sources \citep[for example,][]{lekien, mezic,hall2012}, as well the spatial and statistical
distribution of large scale inhomogeneities \citep[for example,][]{abh}.
The coupling of advective mixing with sinks and sources (other than diffusion) has also proved useful in a geophysical context.
For example, advection-linear damping \citep{chert}, to understand the patchiness of biogeochemical tracers in the ocean with differing lifetimes \citep{Amala1}, advection-reaction-diffusion \citep{neufeld1}, to elucidate the formation and sustenance of plankton blooms \citep{hgl} and
advection-condensation \citep{ray-ac}, to probe the large-scale distribution of water vapor in the atmosphere \citep{sukhatmeYoung2011}. In fact, advection-reaction models have also found use in extraplanetary scenarios, such as understanding seasonal variations in atmospheric composition \citep{titan}.
		
In a two-dimensional (2D) setting, it is well known that 
even relatively simple time dependent flows can lead to complicated tracer
patterns \citep{Aref}. 
An up-to-date review of chaotic mixing, and more broadly mixing implied by multiscale flows, can be found in \cite{Aref-new}, and 
an overview of applications and methods in an oceanographic context can be found in \cite{Prants}. Along with conventional measures such as relative dispersion, regular and anomalous diffusion, Lagrangian tools from dynamical systems such as Finite Time Lyapunov Exponents (FTLEs) and Finite Size Lyapunov Exponents (FSLEs) have proved useful in quantifying stirring via 
simple mixing protocols as well as multiscale turbulent flows. Examples in an oceanic context include, uncovering the mechanisms underlying inter-gyre mixing \citep{Poje-Haller,Wiggins}, transport across jets \citep{Samel},
localized stirring in ocean basins such as the Adriatic \citep{Lacorata}, Tasman \citep{waugh1} and Mediterranean Sea \citep{garcia2007dispersion}, to elucidate the non-uniform nature of surface mixing \citep{waugh2008stirring}, identification of mesoscale eddies \citep{BV-2008} and relative dispersion \citep{Corr}, in the global oceans. In addition to quantifying rates of mixing, these tools also allow for the identification of kinematic transport barriers, i.e., transient structures that inhibit global mixing in geophysical flows \citep{Boff}.
		
In the present work, we bring some of these tools to bear on intraseasonal mixing by geostrophic currents in the Bay of Bengal (BoB) which is a triangular
basin spanning 5$^\circ$-22$^\circ$N and 80$^\circ$-100$^\circ$E, centered around 15$^\circ$N. Apart from an influence on biological activity \citep[for example, evolution of plankton blooms,][]{Vinay-bloom} and better
understanding the dispersal of contaminants [for example, the February 2017 oil spill near Chennai\footnote{\url{https://www.imarest.org/themarineprofessional/item/3104-20-tonnes-of-oil-spilled-in-bay-of-bengal}}], the Bay is an exciting playground with a myriad of seasonal and intraseasonal features. Specifically, the surface flow in the Bay is marked by seasonal features that include an intensified western boundary current that flows northward (equatorward) before (after) the summer monsoon and relatively steady eddies off the eastern coast of India during the monsoon itself \citep{vinayachandran1996forcing,schott2001monsoon}. In addition, altimetry data suggests that the Bay has significant intraseasonal variability in surface geostrophic currents \citep{cheng2013intraseasonal}. In the midst of this activity, another interesting aspect of the Bay is that it is an open basin. In particular, the Bay is connected to the Equatorial Indian Ocean on the southern side and also receives a large amount of fresh water from various river mouths in the northern portion. Much of this river inflow is in the post monsoonal period and comes from Ganga-Brahmaputra and Irrawaddy river basins \citep{papa2012ganga,chait}. In fact, this inflow and its transport is clearly seen in measurements of salinity as well as in numerical simulations \citep{sengupta2016near,Deb1,Akil}, and its dispersal plays an important role in the surface salinity budget of the Bay \citep{raghu}. Taken together, this makes for an interesting dynamical setting in which to assess and quantify the
mixing and dispersal of passive tracers.
		
The outline of this manuscript is as follows: in Section 2, we describe the data used in this study. Section 3 provides an overview of the geostrophic flow from physical and spectral points of view. This gives us a feel for the flow that is responsible for the intraseasonal advection of the passive tracers. Beginning with the advection of latitudinal and longitudinal stripes, in Section 4, we describe and compute FTLEs, Relative Dispersion (RD) and FSLEs. Seasonal mean maps and histograms of these measures are presented so as
to quantify rates and scales, as well as the non-uniformity of chaotic mixing in the Bay. The FSLEs are also used to estimate finite scale diffusion 
coefficients as well as region dependent large scale eddy diffusivities in the Bay. Finally, we use these tools (FTLE \& FSLE) to examine a particular mixing event, specifically, the shielding of postmonsoonal fresh water in October-November 2015 by a persistent eddy in the northern Bay. A discussion and summary of results concludes the paper.
		
\section{Data}
		
The Ssalto Duacs/gridded multimission altimeter products, which are a part of AVISO project (\url{http://www.aviso.altimetry.fr/en/home.html}), have been used in this study. Specifically, we use MADT-H-UV (Maps of Absolute Dynamic Topography \& Absolute Geostrophic Velocities) datasets with a spatial resolution of $0.25^{\circ} \!\times \!0.25^{\circ}$, covering $5^{\circ}$N to $24^{\circ}$N and $80^{\circ}$E to $100^{\circ}$E on a Cartesian grid. To verify the robustness of our results we have used multiple years of data, specifically, 2010-2013. For salinity, we have used the Level-3 SMAP SSS version-3 data set produced by the Jet Propulsion Laboratory (\url{https://podaac.jpl.nasa.gov/dataset/SMAP_RSS_L3_SSS_SMI_8DAY-RUNNINGMEAN_V2?ids=Collections&values=SMAP-SSS}) at 0.25$^\circ$ horizontal resolution and 8 day running average time window from 31 March 2015 to 31 December 2015. 
		
\begin{figure}[h]
	\centering
	\includegraphics[scale = 0.7]{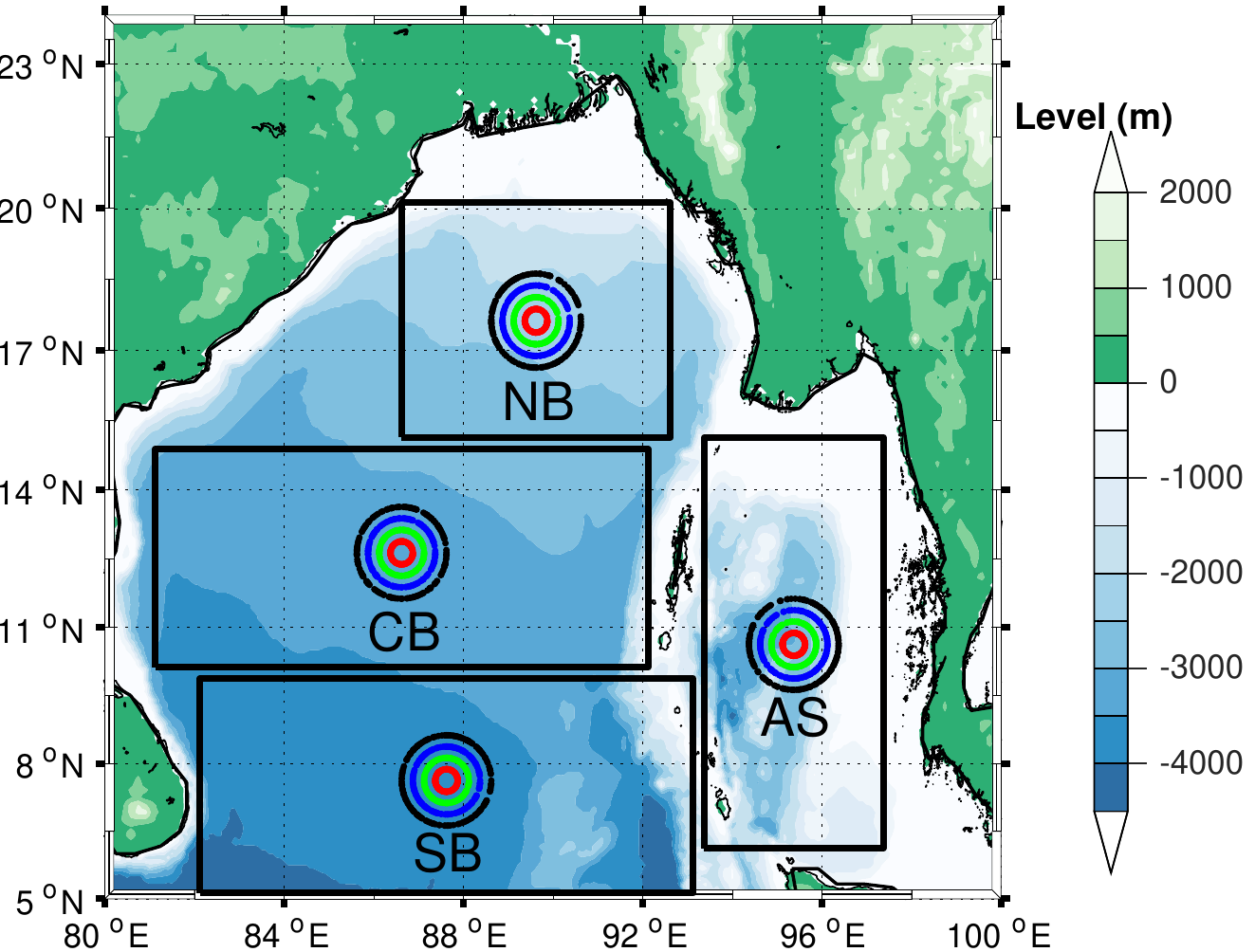}
	\caption{The Bay of Bengal and subregions [Northern Bay (NB), Central Bay (CB), Southern Bay (SB) and Andaman Sea (AS)].}
	\label{fig1}
\end{figure}

Based on its general circulation \citep{potemra1991seasonal}, the Bay has been divided into four different regions in this study --- Northern Bay (NB $\rightarrow$ $15.125^\circ$N-$20.125^\circ$N, $86.625^\circ$E-$92.625^\circ$E), Central Bay (CB $\rightarrow$ $10.125^\circ$N-$14.875^\circ$N, $81.125^\circ$E-$92.125^\circ$E), Southern Bay (SB $\rightarrow$ $5.125^\circ$N-$9.875^\circ$N, $82.125^\circ$E-$93.125^\circ$E) and Andaman Sea (AS $\rightarrow$ $6.125^\circ$N-$15.125^\circ$N, $93.375^\circ$E-$97.375^\circ$E). 
The geographic locations of these boxes are shown in Figure \ref{fig1}. This division, and expected heterogeneity in mixing, is based on the general circulation of the Bay \citep{potemra1991seasonal}, the natural partitioning provided by the Andaman Islands and the inflow of fresh water that distinguishes the northern and southern portions of the Bay. 
		
\section{Circulation in Bay of Bengal}
		
Given its importance to the regional climate, the seasonal circulation of the BoB has been studied quite extensively \citep[see for example,][]{potemra1991seasonal,vinayachandran1996forcing,schott2001monsoon}, and we refer the readers to the aforementioned papers for details. Here, we provide an overview of the subseasonal surface geostrophic circulation to get a feel for the flow that is responsible for advection in the subsequent passive tracer mixing experiments.
		
\subsection{Physical space characterization}
		
On intraseasonal timescales, in the context of geostrophic surface flows, 
the co-existence of Rossby waves and (nonlinear) eddies is well established throughout the world's oceans \citep[see for example,][]{chelton2007}. In the BoB too, a rich interplay of Rossby waves and eddies has been noted in numerous studies 
\citep[for example,][]{shankar1996dynamics,babu2003circulation,kurien2010mesoscale,chen2012features,nuncio2012life}. 
It has been suggested that the west coast of the Bay is a critical region for eddy-mean flow interaction with significant baroclinic instability and the  production of eddies \citep{kurien2010mesoscale,chen2012features,nuncio2012life}.
Further, eddies in the BoB have different propagation characteristics in different regions. For example, eddies in the northern and southern Bay (i.e., north of 15$^{\circ}$N and south of 10$^{\circ}$N) propagate in southwestward and northwestward directions, respectively. While those in the central Bay tend to move in along the same latitude in a westward direction \citep{chen2012features}.
		
\begin{figure*}[h!]
	\begin{center}
	%
		\subfigure[]{%
		\label{fig 2.1}
		\includegraphics[width=0.45\textwidth]{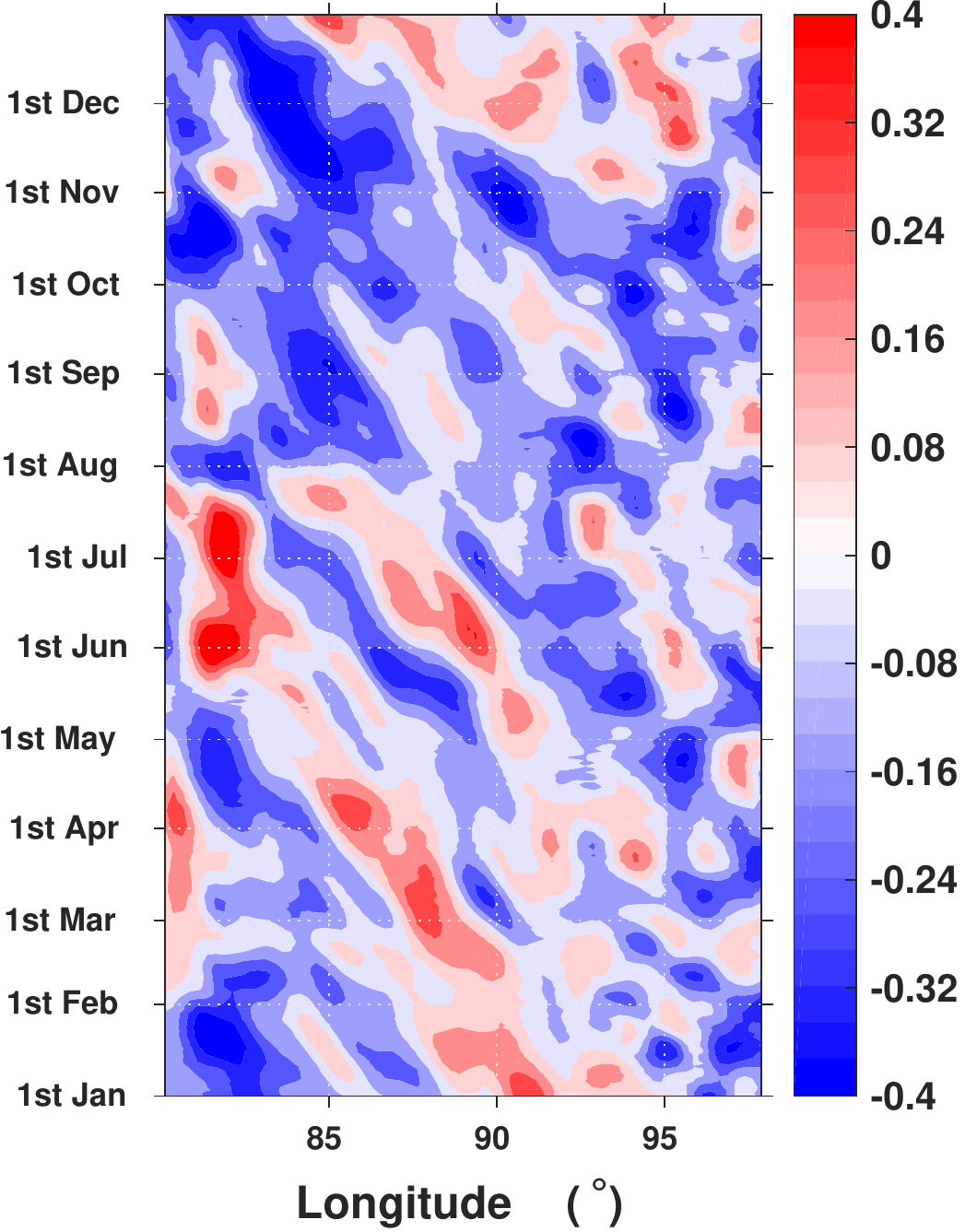}             
		}%
		\subfigure[]{%
		\label{fig 2.2}
		\includegraphics[width=0.45\textwidth]{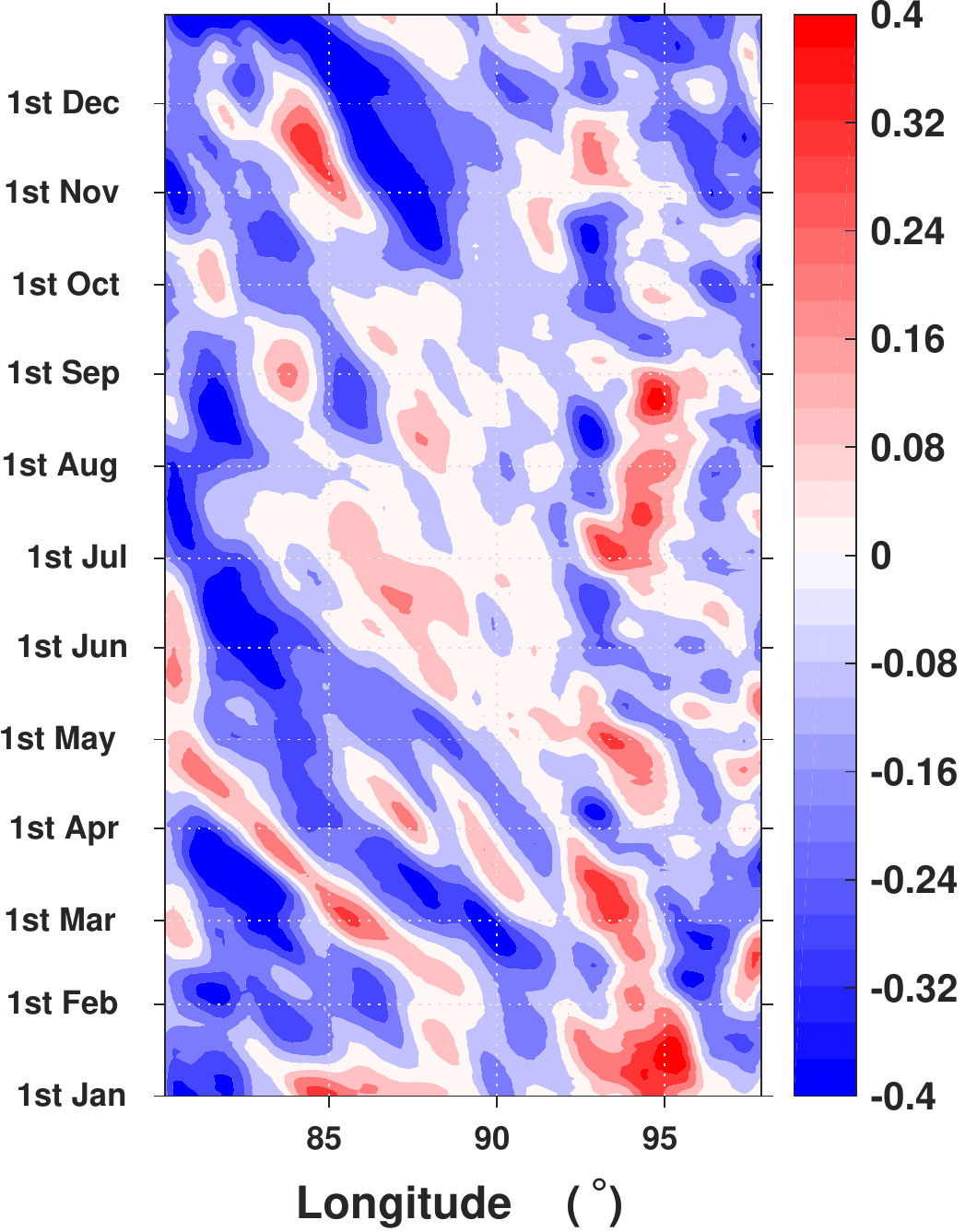}
		}\\ 
	  	\subfigure[]{%
		\label{fig 2.3}
		\includegraphics[width=0.45\textwidth]{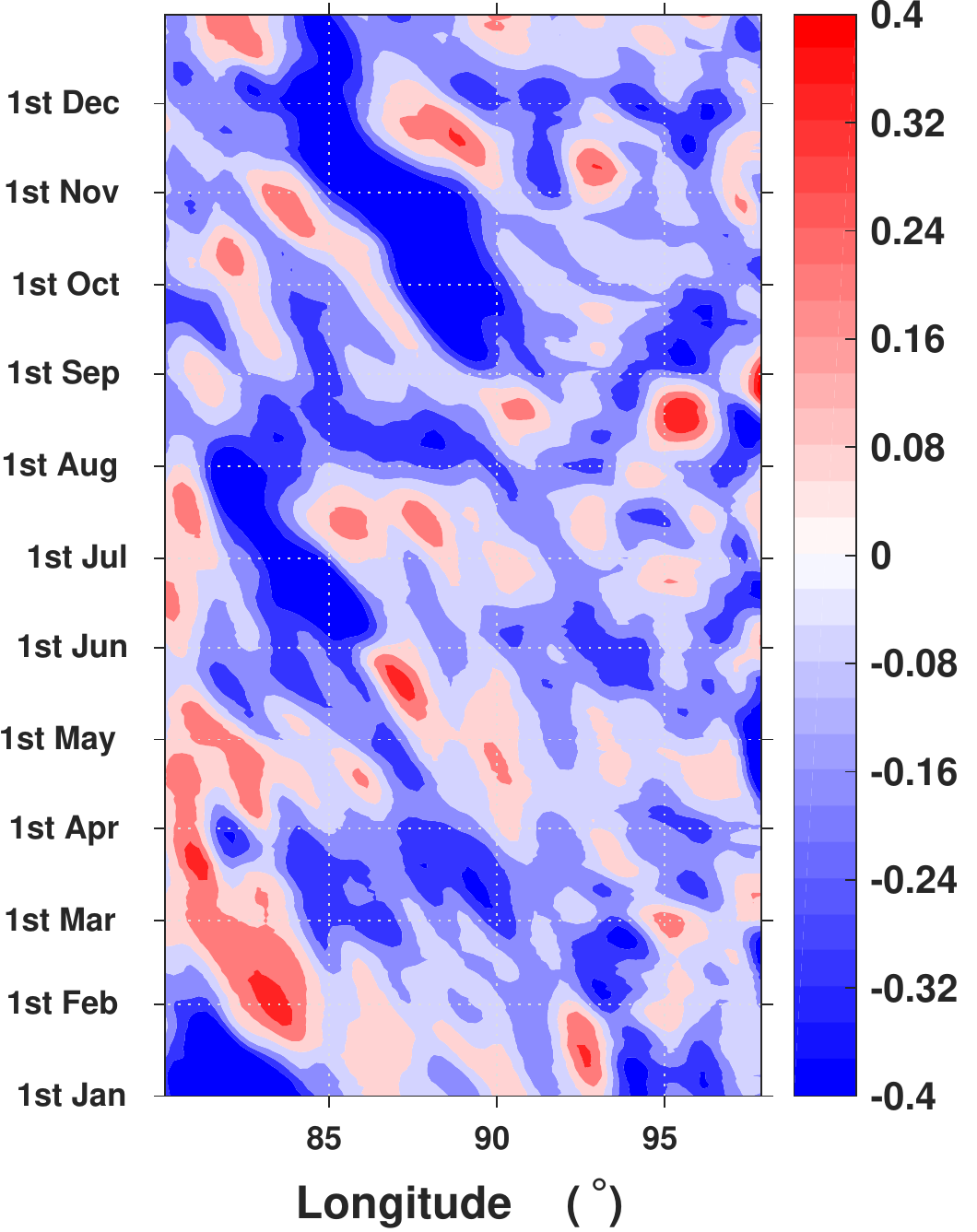}
		}%
		\subfigure[]{%
		\label{fig 2.4}
		\includegraphics[width=0.45\textwidth]{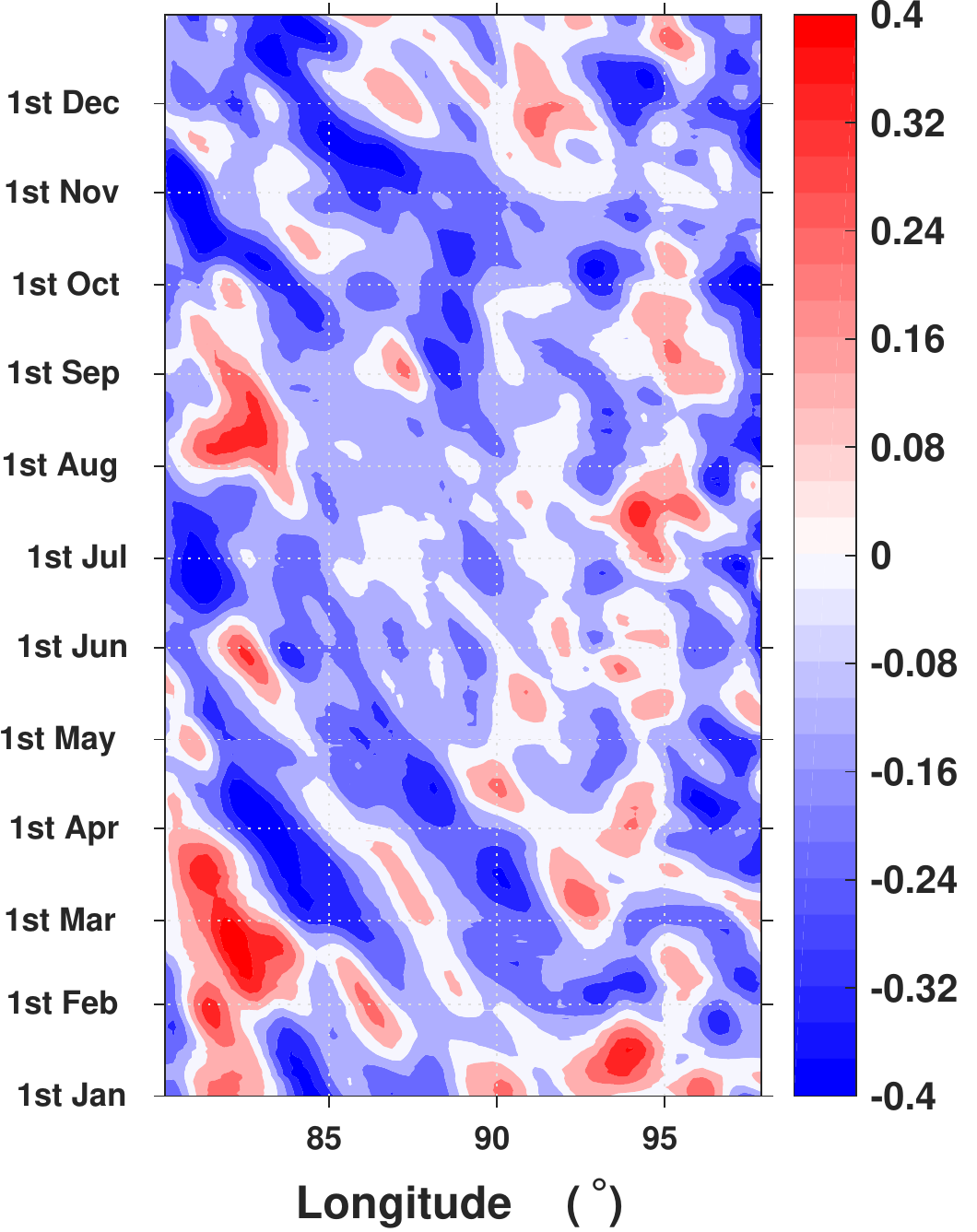}
		}\\
				
		\end{center}
		\caption{Panels (a), (b), (c) and (d) show Hovm{\"o}ller diagrams of the zonal geostrophic velocity in 2010, 2011, 2012 and 2013, respectively. 
		The plots are from a longitude of 80$^{\circ}$E to 100$^{\circ}$E in 
		the BoB, the cross section is taken near the widest portion of the central Bay at 12.125$^{\circ}$N }%
		\label{fig2}
\end{figure*}

We begin with Hovm{\"o}ller plots with respect to longitude of the zonal geostrophic velocity in different seasons of all the four years (meridional velocities show essentially similar features). The longitude spans $80^{\circ}$E to 100$^{\circ}$E, with the latitude being fixed at the Central Bay (this is the widest portion of the Bay, hence allows for a clear observation of the east-west movement of disturbances), specifically, $12.125^{\circ}$N. As seen in Figure \ref{fig2}, throughout the year we observe southeast to northwest tilted coherent structures. The tilts are fairly steady across the years and yield a westward phase speed of approximately 8-12 $cm/s$, 
consistent with prior estimates from the central BoB \citep{killworth1997speed,chelton2007,chen2012features}. Interestingly, these cohesive southeast to northwest tilting structures, or wave packets, also appear to exhibit episodic eastward migration with systematic positive (red) and negative (blue) anomalies. Specific examples are : March to July between 85$^\circ$E to 95$^\circ$E in 2010, February to May around 85$^\circ$E and October to December between 85$^\circ$E and 90$^\circ$E in 2011, October to December between 85$^\circ$E to 90$^\circ$E in 2012 and February to May around 85$^\circ$E in 2013. These suggest a small eastward group velocity associated with these wave packets. 
We have also constructed Hovm{\"o}ller plots of the zonal velocity with respect to latitude (not shown), these showed a fairly mixed behavior with sporadic instances of northward and southward tilts. By and large, the movement of disturbances in the East-West direction in the Bay is more pronounced and systematic as compared to North-South migration.
		
\begin{figure*}
	\begin{center}
		\subfigure[]{
		\label{fig 3.1}
		\includegraphics[scale=0.8]{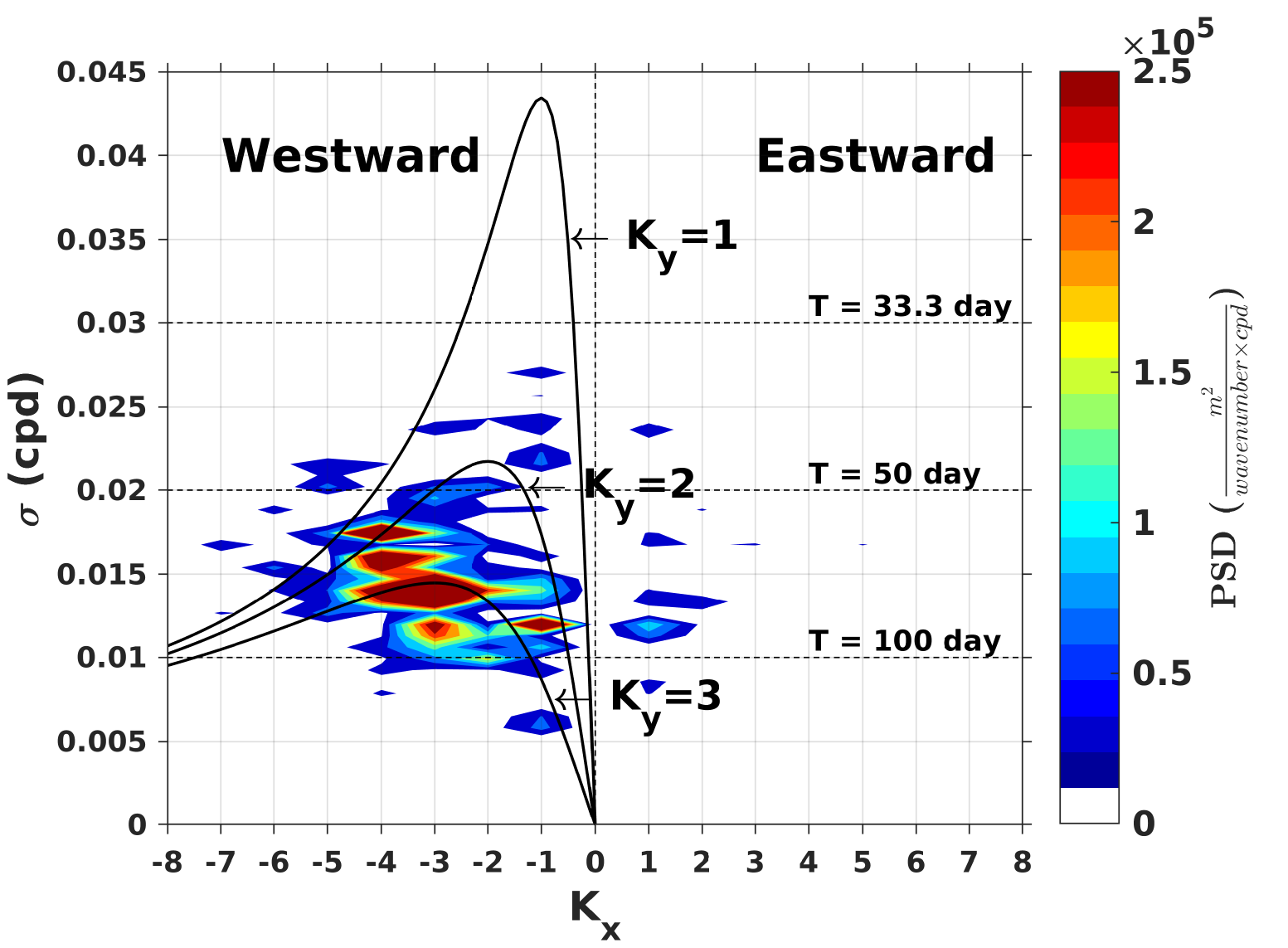}}
		\subfigure[]{
		\label{fig 3.2}
		\includegraphics[scale=0.8]{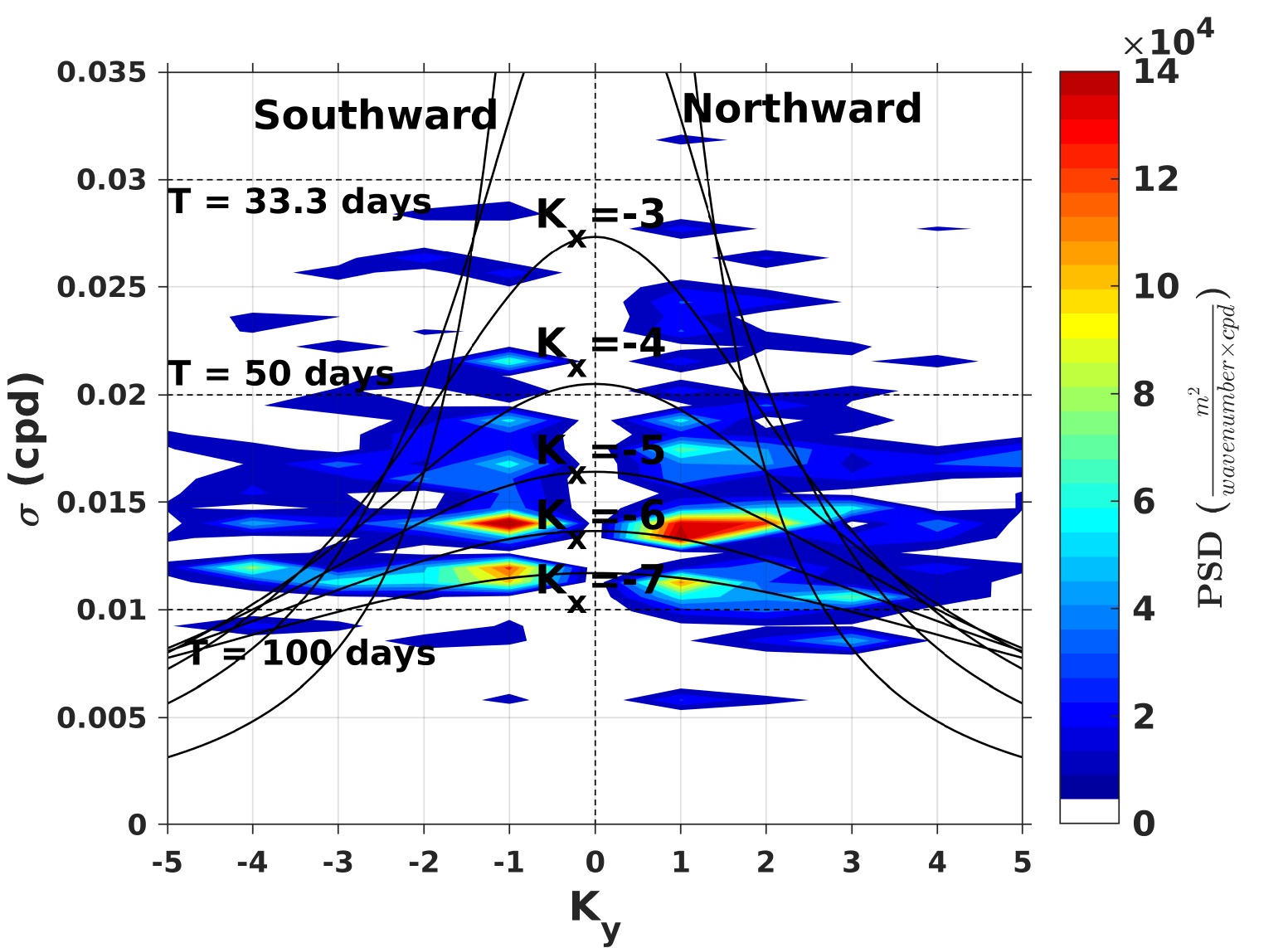}}
		\caption{Wavenumber-frequency plots of SLA. Panels (a) and (b) show $\sigma-K_x$, averaged over 14$^\circ$N-15$^\circ$N and $\sigma-K_y$, 
		averaged over 89.5$^\circ$E-90.5$^\circ$E, respectively. Solid black lines are theoretical linear Rossby wave dispersion curves.}
	\end{center}
		\label{fig3}
\end{figure*}
		
\subsection{Spectral space characterization}
		
To get a feel for the scale and time period of these disturbances we move to spectral space and compute wavenumber-frequency diagrams of the sea level anomaly (SLA) constructed from MADT-H. The climatological mean has been removed from the data to construct the SLA and then it is subjected to a 30-120 day band pass Lanczos filtering, with number of coefficient equals to 100. The results presented are averaged over a latitude band from 14.125$^\circ$N to 15.125$^\circ$N (through the central and widest longitudinal extent of the Bay, and for consistency with the Hovm{\"o}ller plots). As seen in Figure \ref{fig 3.1}, we note strong spectral peaks spread over $|k_x| \approx 1-5$ for $k_x < 0$, with time
periods ranging from about 35 to 100 $days$. The three solid curves shown in Figure \ref{fig 3.1} correspond to the dispersion relations of baroclinic Rossby modes given by,
		
\begin{equation}\label{eq:1}
	\sigma = -\frac{\beta k_x}{{k_x}^2+{k_y}^2 + \frac{1}{L_R^2}}, \textrm{for } k_y = 1, 2, 3. 
\end{equation}
		
Here, $k_x$ and $k_y$ are the zonal and meridional wavenumber and $L_R$ is the Rossby radius of deformation. The wavenumbers have been normalized by the length of the BoB which is equal to 1800 $km$ (72 $\times$ 1/4 $^\circ$) at 15$^\circ$N. 
The typical value of $L_R$, lies between 50-100 $km$ over the latitudes spanned by the Bay \citep{chelton1998geographical}. In fact, following \cite{stammer1997global}, we estimated an ``eddy length scale" from zero crossing of autocorrelation function of the SSH. This estimate (not shown) is somewhat larger than $L_R$, and varies from 125-165 $km$ over 18$^\circ$N and 5$^\circ$N. The peaks in the left half of Figure \ref{fig 3.1} correspond to westward phase speeds, and they appear to be guided by Equation \ref{eq:1}. This suggests that these disturbances with westward phase speed are related to baroclinic Rossby waves. Further, the position of the maxima in these diagrams show that there is significant power at scales just smaller than the local
deformation scale (maxima of the dispersion curves), thus supporting episodic eastward group velocities noted in the longitudinal Hovm{\"o}ller plots in Figure \ref{fig2}. 
		
We also computed a dispersion diagram of frequency vs meridional wavenumber, and this is shown in Figure \ref{fig 3.2}. The spectra are averaged over a longitudinal band of 89.625$^\circ$E-90.625$^\circ$E (where we have the largest latitudinal extent of the Bay). The meridional wavenumber has been normalized by the longitudinal cross-section of BoB which is equal to 1675 $km$ (67 $\!\times\!$ 1/4 $^\circ$) at 90$^\circ$E. We note that the most of the power is distributed between $|k_x| \approx 1-5$, with a maximum between $1<|k_y|<2$. Also, power appears to be distributed  asymmetrically, higher in the northward direction than southwards. Both the wavenumber-frequency plots, $\sigma-k_x$ and $\sigma-k_y$, have largest power in the temporal band of 60-90 days.
		
\begin{figure*}[h]
	\begin{center}
		\subfigure[]{
			\label{fig 4.1}
			\includegraphics[scale=0.5]{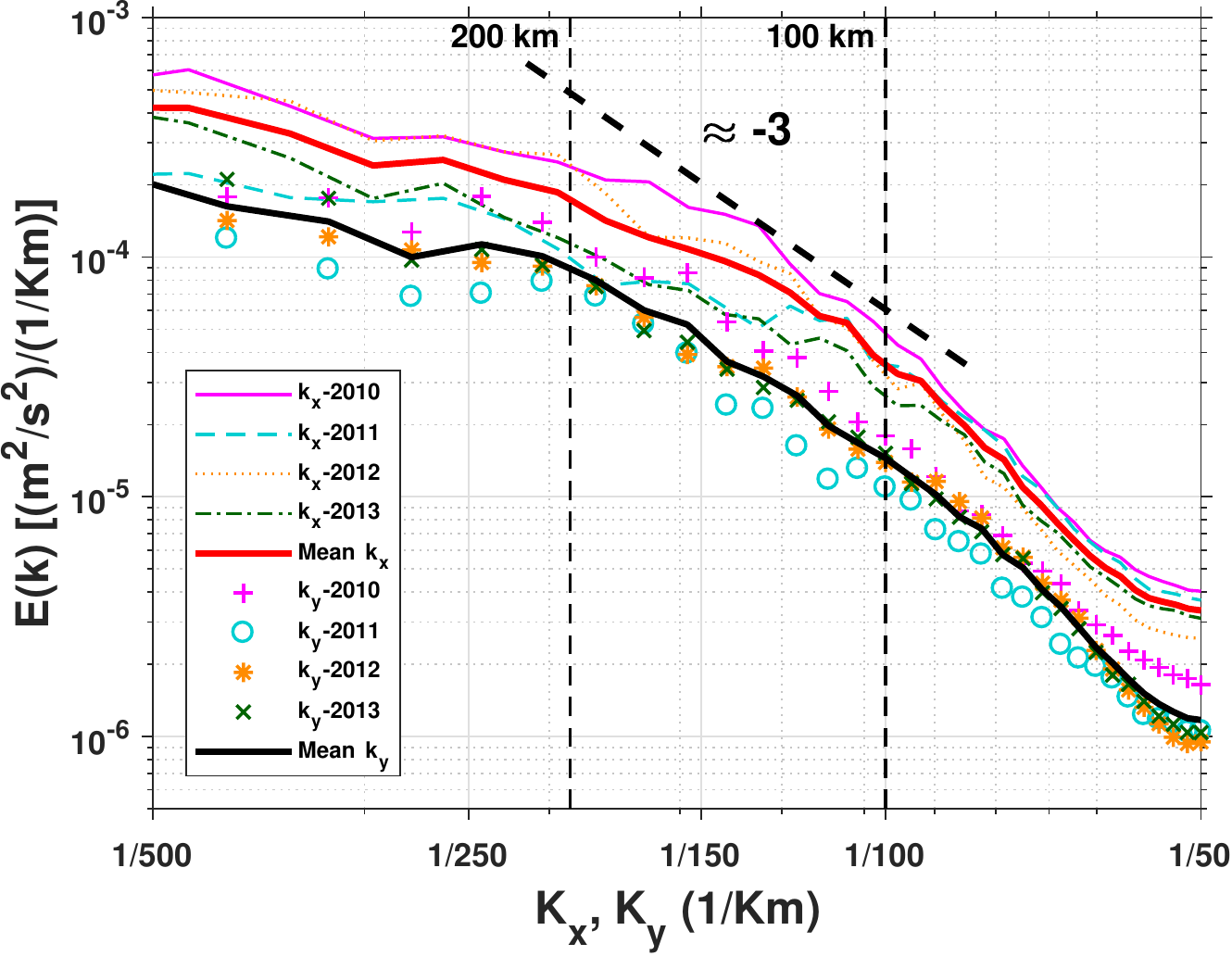}}
		\subfigure[]{
			\label{fig 4.2}
			\includegraphics[scale=0.5]{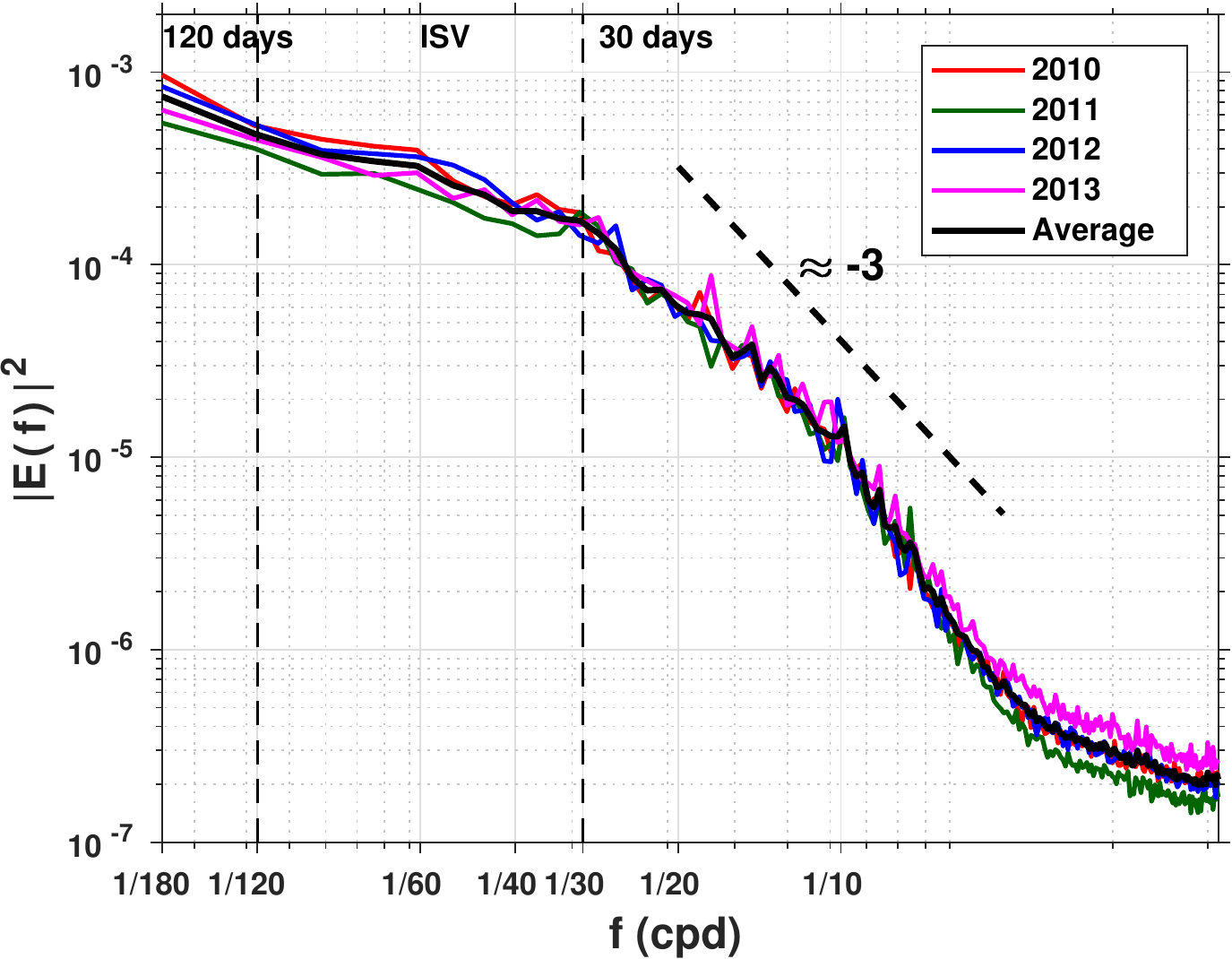}}\\
		\subfigure[]{
			\label{fig 4.3}
		\includegraphics[scale=0.5]{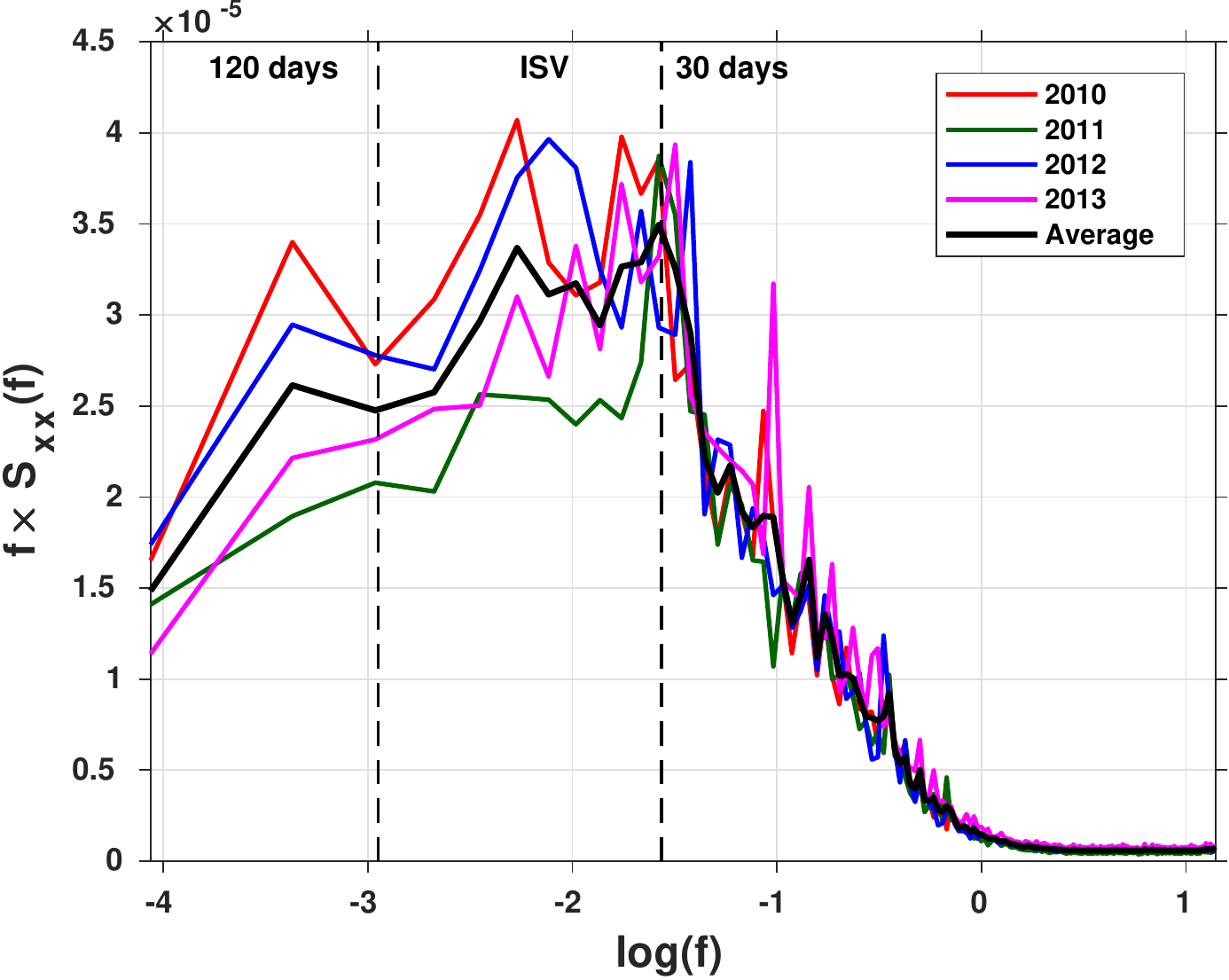}}
		\caption{Kinetic energy spectra. Panel (a) shows meridional and zonal wavenumber spectra averaged over 11.125$^\circ$N-12.125$^\circ$N and 
		89.625$^\circ$E-90.625$^\circ$E, respectively. Panels (b) and (c) contain temporal spectra vs $f$ (frequency) in log-log scale and variance preserving form, respectively. Spectra are estimated at each grid point and then averaged.}
		\label{fig4}
	\end{center}
\end{figure*}

Finally, power spectra are examined to get an idea of the energy associated with the different length and time scales in the geostrophic velocity field.
Specifically, we compute kinetic energy spectra and average over 11.125$^\circ$N to 12.125$^\circ$N which resolves largest zonal scales in the Bay. In a similar manner, spectra for meridional wavenumber have been averaged over 89.675$^\circ$E to 90.675$^\circ$E (again, these particular longitudes are chosen to resolve the largest meridional scales in the Bay). The slopes of the spectra, as seen in the first panel of Figure \ref{fig 4.1}, are all close to a $-3$ power-law between approximately 100 $km$ and 250 $km$ in both the zonal and meridional directions. This is agreement with a
forward enstrophy cascading regime of surface geostrophic currents from about 200 $km$ to 100 $km$ in the global open oceans \citep[see, for example, the detailed discussion in][]{khatri}. The temporal spectrum is estimated by calculating the spectrum at each grid point and then averaging. This is done for each year and the results are presented in log-log and variance preserving form in the second and third panels of Figure \ref{fig4}, respectively. The variance preserving form, in agreement with the wavenumber-frequency plots, shows relatively isolated peaks at scales longer than 30 $~days$. Further, the log-log plot shows signs of an approximate power-law, with a $-3$ exponent, for time scales that range from about 10-30 $days$. The match in
temporal and spatial spectral exponents suggests that Taylor's hypothesis appears to hold at these scales; specifically, using an annual average speed of 0.2 $m/s$, a spatial scale of 200 $km$ maps to approximately 10 $days$ \citep[see, for example,]
[]{wf,ss}. Note that, given the approximate ten day frequency of repeat satellite passes in AVISO, periods below two weeks suffer from aliasing, and the spectra at these smaller timescales are likely to be unreliable \citep{Arbic}. Taken together, the spatial and temporal spectra suggest an uninterrupted distribution of power across mesoscales and from intraseasonal to weekly time scales, respectively.
		
\section{Mixing of Passive Tracers}
		
As demonstrated, geostrophic flow in the BoB has a multiscale character in both space and time. Specifically, there is strong seasonal dependence of the surface flow 
\citep[for example, current disruption and reversal,][]{vinayachandran1996forcing,schott2001monsoon}, along with significant subseasonal variability consisting of geostrophically balanced disturbances that 
exhibit predominantly westward phase speeds and align quite well with the dispersion curves for baroclinic Rossby waves. All together, the flow provides a rich playground for the mixing of passive fields. As it happens, aperiodic Rossby waves by themselves have been examined in detail as idealized models of chaotic mixing \citep[see, for example,][]{Ray1}. The principal tool used in these mixing calculations is the Lagrangian advection of parcels. This is done using a Runge Kutta fourth order (RK4) scheme. Further, given that the data is on a fixed grid, the flow has been interpolated by a bilinear interpolation scheme. 
		
As our measures of mixing are Lagrangian in nature, it is worth identifying the limits imposed on the calculations due to the resolution of the altimeter data. In general, as discussed by \cite{Bart}, coarse velocity field data performs satisfactorily with regard to passive advection when its kinetic energy spectrum follows a $-3$ power-law. Of course, finer scale data can improve quantification of Lagrangian measures 
\citep{BV-2010}. As seen in Figure \ref{fig 4.1}, currents in the Bay follow this scaling over a range of approximately 100 to 250 $km$. But, the situation
is complicated at smaller scales. Specifically, at the ocean's surface, scales below approximately 100 $km$ (depending on the region in consideration) have a significant contribution from the divergent component of the flow \citep{capet2008mesoscale,BCF,B2,qiuetal}, and spectra at these scales smaller also show signs of flattening to shallower power-laws \citep{callies}. 
Not only is the altimeter derived geostrophic data attenuated by filtering below scales of approximately 100 $km$ \citep[see, for example,][]{arbic2013eddy,altimeter}, it is unlikely to be a dominant contributor to the actual surface currents. Thus, caution must be exercised in interpreting Lagrangian measures computed from purely geostrophic data below a scale of approximately 100 $km$.

\subsection{A Sign of Chaos}
		
\begin{figure*}[h]
	\begin{center}
		\subfigure[]{
		\label{fig 5.1}
		\includegraphics[width=14cm,height=9.5cm]{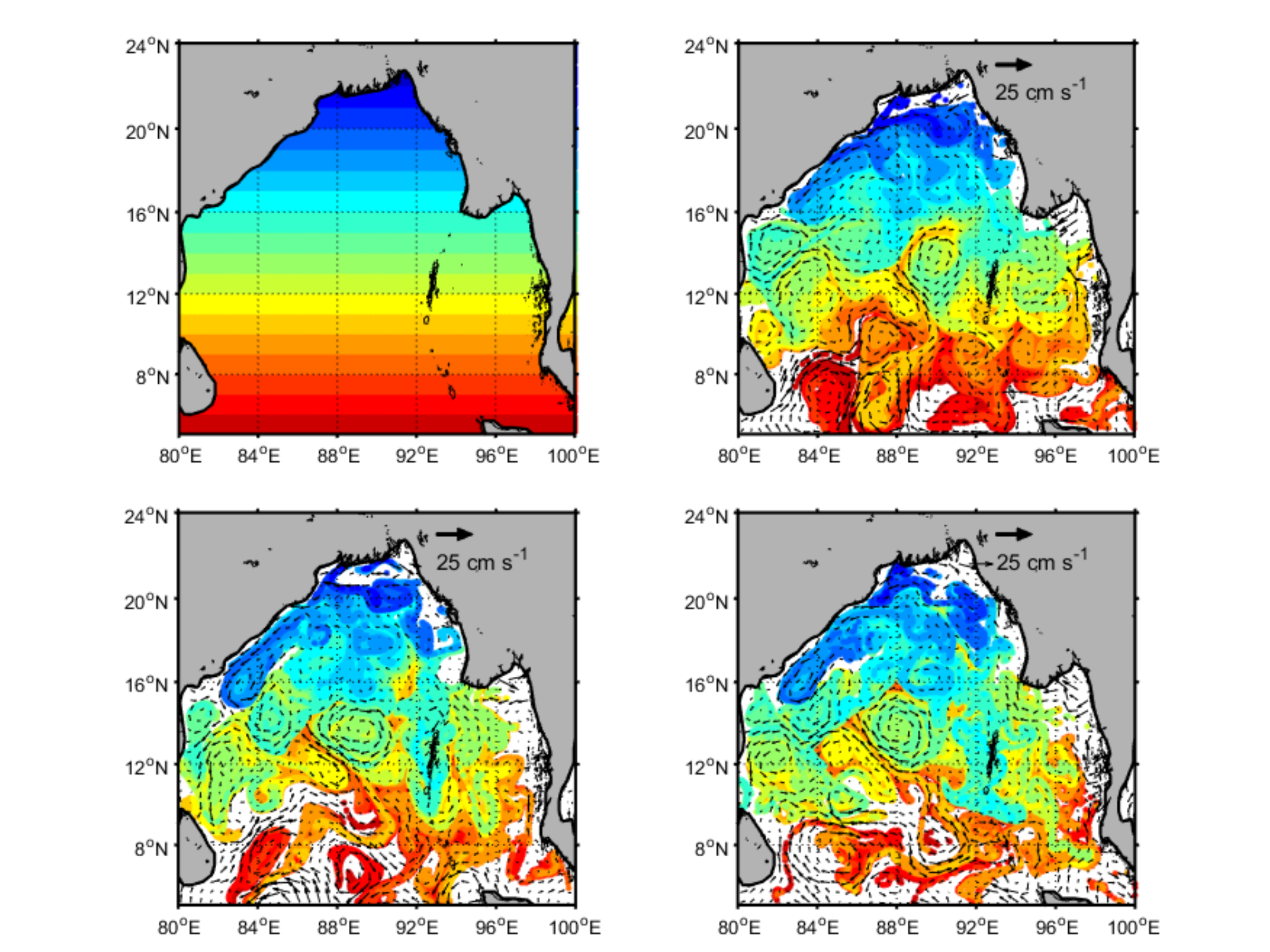}}
		\subfigure[]{
		\label{fig 5.2}
		\includegraphics[width=14cm,height=9.5cm]{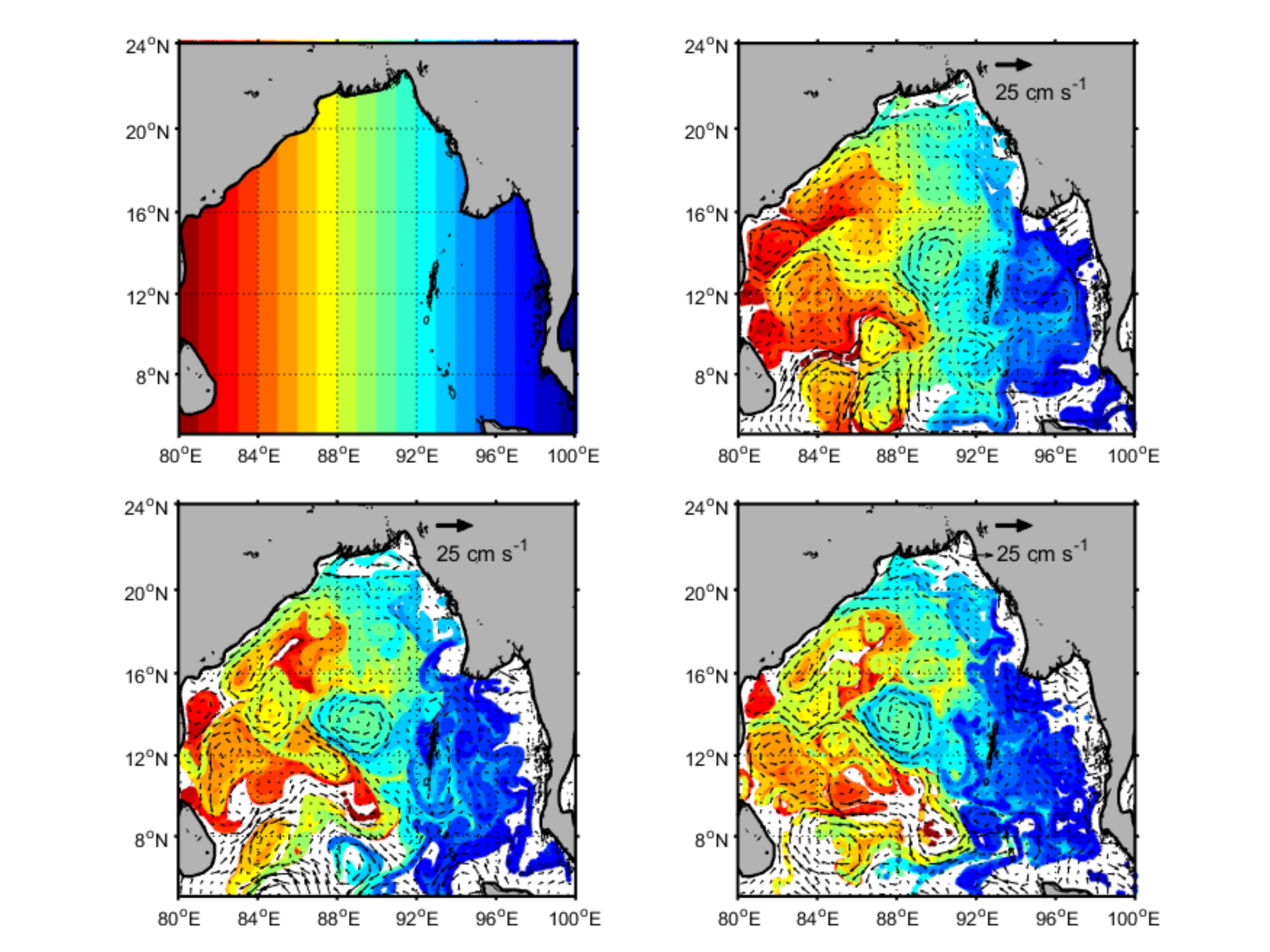}
		}	
		\caption{The stirring of latitudinal (first four panels) and longitudinal (last four panels) bands by geostrophic currents from September to October 2012. Snapshots are shown every two weeks for six week long advection.}%
		\label{fig5}
		\end{center}
\end{figure*}

We begin with a simple numerical experiment where the BoB is divided into zonal and longitudinal sections. Each band is approximately 2$^\circ$ wide and is identified by a separate color as seen in the first two panels of Figure \ref{fig5}.
Parcels in each band retain their color and are advected for about six weeks. Such experiments give a basic feel for the mixing processes at work \citep[see, for example,][]{Ray}. Snapshots of the scalar field are shown every two weeks in Figure \ref{fig5}. We notice that within the first two weeks, the bands are distorted and form extended filaments and whorls. In fact, the boundaries between the different colors evolve towards highly complicated contours. This process continues with the repeated wrapping around of progressively thinner scalar filaments, i.e., a cascade of 
tracer variance to small scales. Indeed, the geometric picture that emerges is 
that of chaotic advection of a tracer field \citep{ottino}, and by week six it is quite difficult to distinguish between the longitudinal and latitudinal bands. With this qualitative picture in mind, we now proceed to more quantitative measures of mixing in the BoB.
		
\subsection{Finite Time Lyapunov Exponents}
		
A fundamental quantitative characteristic of a chaotic flow is its Lyapunov exponent, this is defined as the exponential rate of separation, averaged over infinite time, of fluid parcels with initial infinitesimal separation \citep{benettin1980lyapunov}.
For practical problems, the limit of time tending to $\infty$ is not feasible, and the notion is generalized to the so-called Finite Time Lyapunov Exponents (FTLE, $\lambda_T$). Specifically, 
		
\begin{equation}\label{eq:2}
	\lambda_T(\tau) = \lim_{r(0) \to 0} \frac{1}{\tau} \log \frac{r(\tau)}{r(0)}.
\end{equation}
		
For non-autonomous flows, the FTLE is essentially a measure of integrated strain along a parcel's trajectory. Here, $\lambda_T(\tau)$ is calculated from the logarithm of the largest eigenvalue of $M^TM$, where $M$ is the deformation matrix obtained
by integrating the Jacobian of the flow along a trajectory \citep[details can be found in,][]{haller2002lagrangian,abraham2002chaotic,waugh1}. A finite difference scheme is employed to compute the Jacobian and its traceless component is used to obtain $M$, which ensures that $\lambda_T(\tau)$ is positive.
		
\begin{figure*}
	\begin{center}
		\includegraphics[width=1\textwidth]{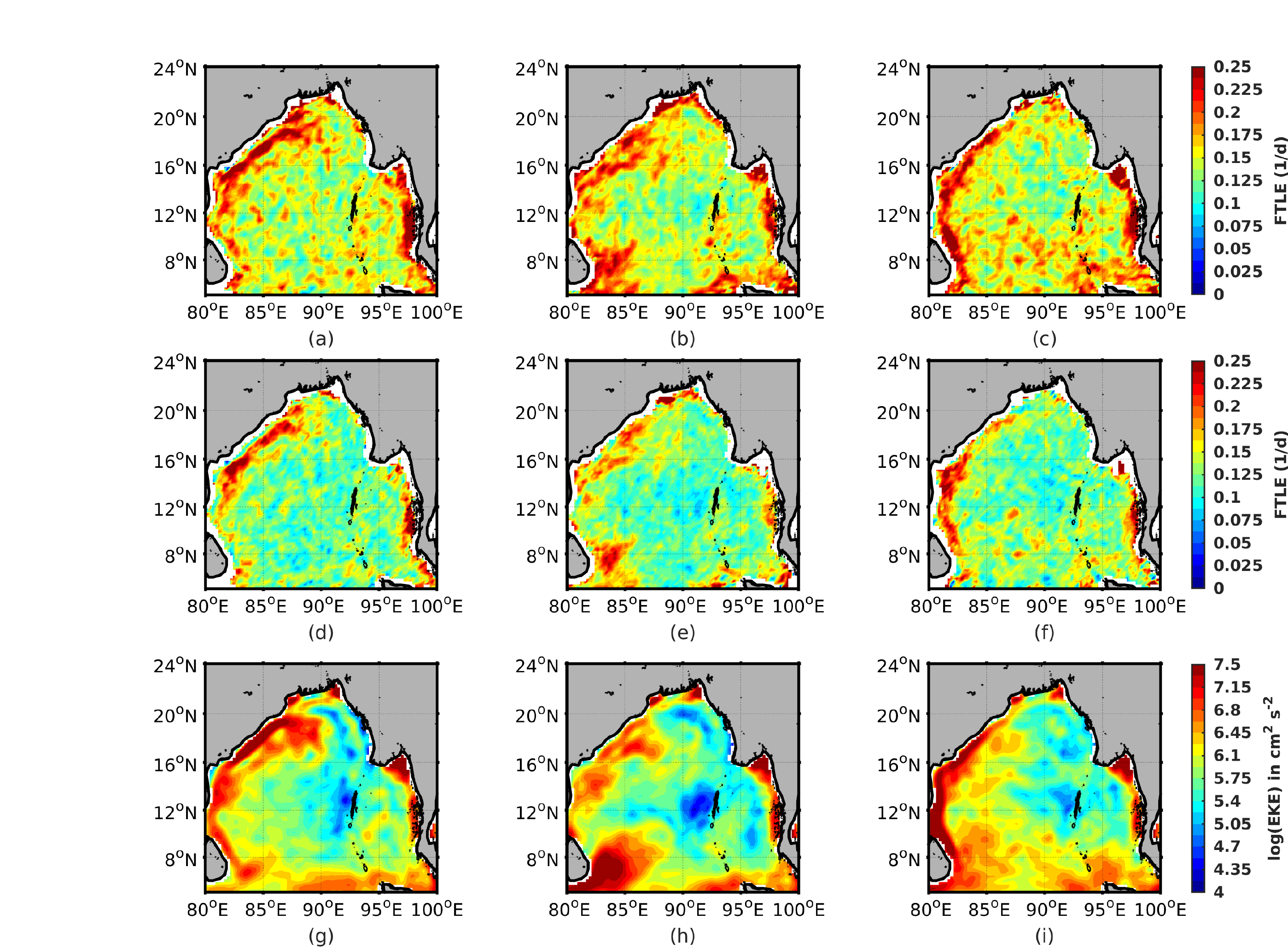}
		\caption{Panels (a) and (d) show the FTLE for 5 and 10 days in FMA, (b) and (e) for JJAS, (c) and (f) for the OND season. Panels (g), (h) and (i) show the corresponding eddy kinetic energy for FMA, JJAS and OND seasons, respectively.}
		\label{fig6}
	\end{center}
\end{figure*}

FTLEs are estimated for $\tau= 5, 10, 15$ and 20 $days$ for all the months of four years. Proceeding seasonally, we have averaged FTLEs for Feb-Mar-Apr (FMA), Jun-Jul-Aug-Sep (JJAS) and Oct-Nov-Dec (OND). Further, the average is for seasons from all four years of data (2010-2013). Figure \ref{fig6} shows the resulting seasonal FTLE maps for $\tau=5$ and 10 $days$. Positive values of the FTLE indicate chaotic behavior over most regions of the Bay. In the premonsoon season (FMA), the FTLEs after 5 and 10 $days$ are high along the western boundary and the northern Bay with values
of approximately 0.15-0.3 $day^{-1}$ and 0.1-0.2 $day^{-1}$, respectively. Interestingly, for $\tau=10$ $days$, the southwest portion of the Bay (near the Andaman Sea) also shows strong chaotic mixing. The picture changes in the monsoon period (JJAS), with higher FTLEs on the western boundary moving systematically
equatorward. Another notable feature of JJAS is the emergence of a localized pocket of high FTLEs off the east coast of Sri Lanka, this enhanced mixing during the monsoon is due to the so-called Sri Lankan dome \citep{Vinay-Yama}. Further, the central regions of the Bay show relatively smaller FTLEs, suggestive of kinematic barriers to basin wide mixing between the eastern and western portions of the Bay. This is consistent with Figure \ref{fig5} where we saw stretching and folding at scales smaller than the basin size. Finally, in OND, the southern part of the Bay lights up with high FTLEs. A close inspection of the maps in OND shows signs of a ring of high FTLEs between 85$^\circ$-90$^\circ$E and 8$^\circ$-12$^\circ$N, a feature sometimes referred to as the BoB dome \citep{Vinay-Yama}. The northwest boundary is now relatively quiescent with low FTLE values. 
Overall, we see a seasonal cycle in mixing with rapid stirring progressively moving from the northern to southern Bay, from pre-monsoonal to post-monsoonal periods, respectively. Further, the eddy kinetic energy (EKE) in the Bay (where eddies are defined as deviations from a climatological 4 year mean) in each season is also shown in the third panel of Figure \ref{fig6}. Clearly, the EKE map aligns quite closely with that of the FTLEs in each season. This has been noted on global \citep{waugh2008stirring}, as well more local scales \citep{waugh1}, in the world's oceans. 
		
\begin{figure*}[h]
			
		\subfigure[]{
		\label{fig 7.1}
		\includegraphics[width=0.4\textwidth]{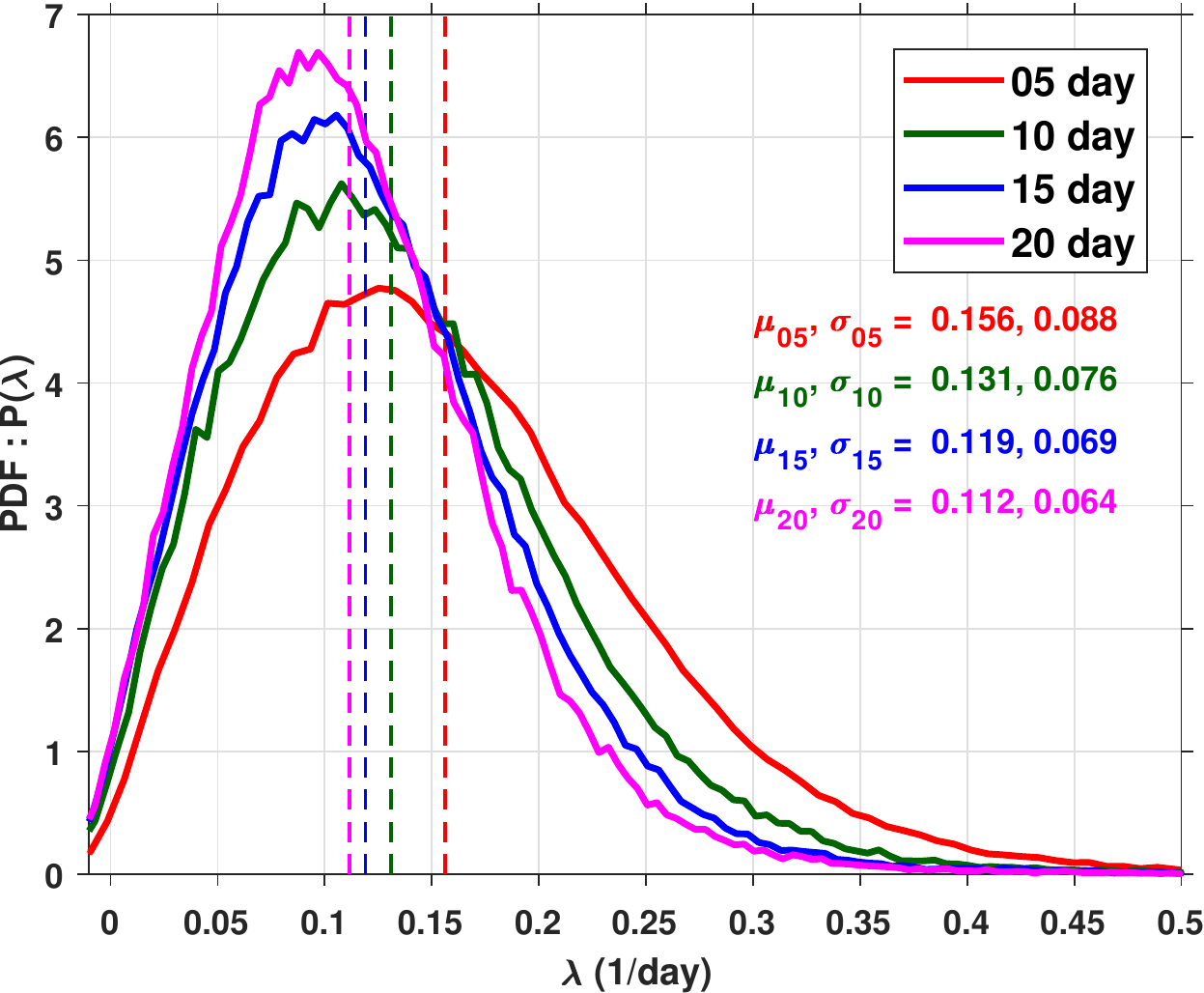}
		}
		\subfigure[]{
		\label{fig 7.2}
		\includegraphics[width=0.4\textwidth]{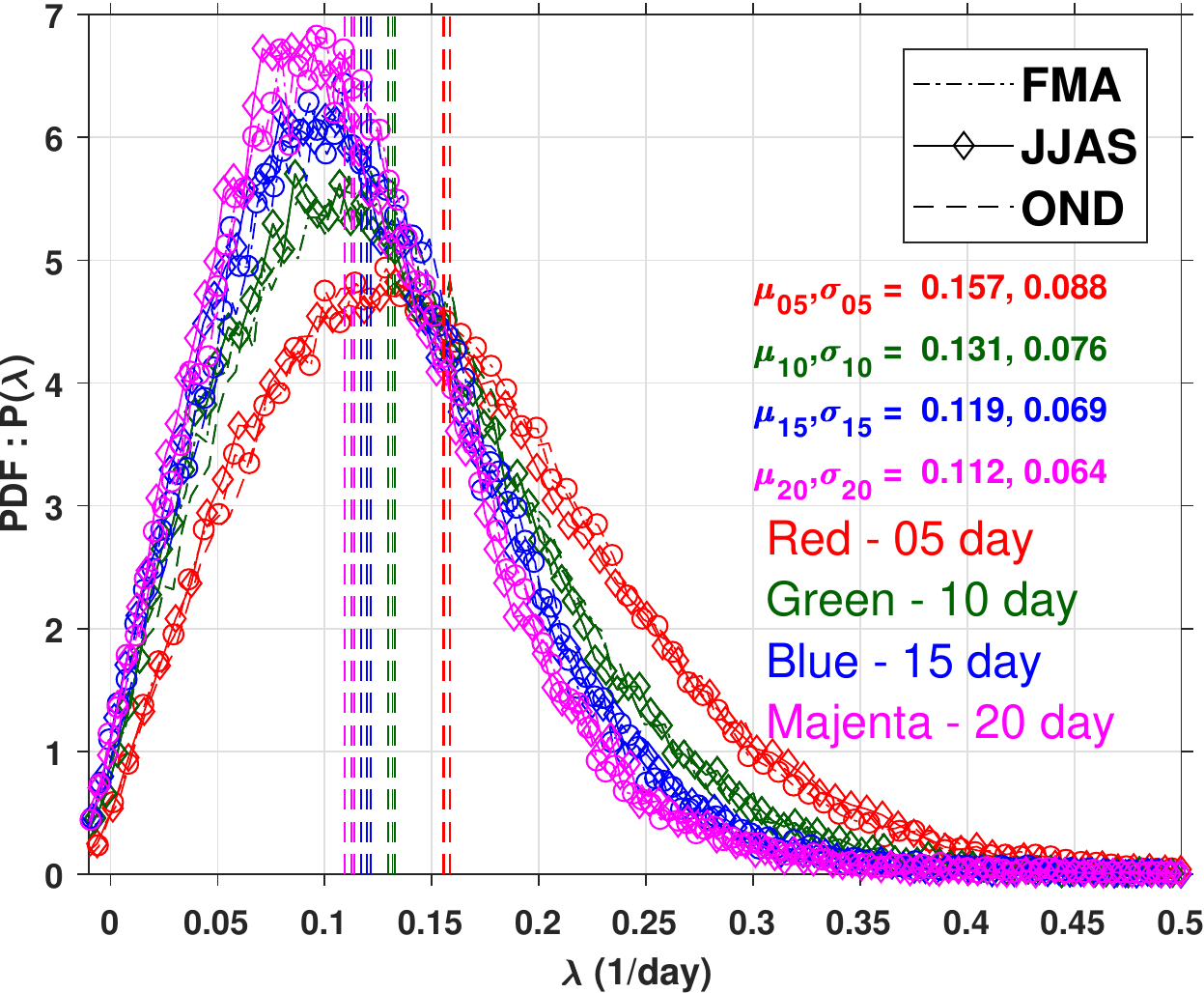}
		}\\
		\subfigure[]{
		\label{fig 7.4}
		\includegraphics[width=0.4\textwidth]{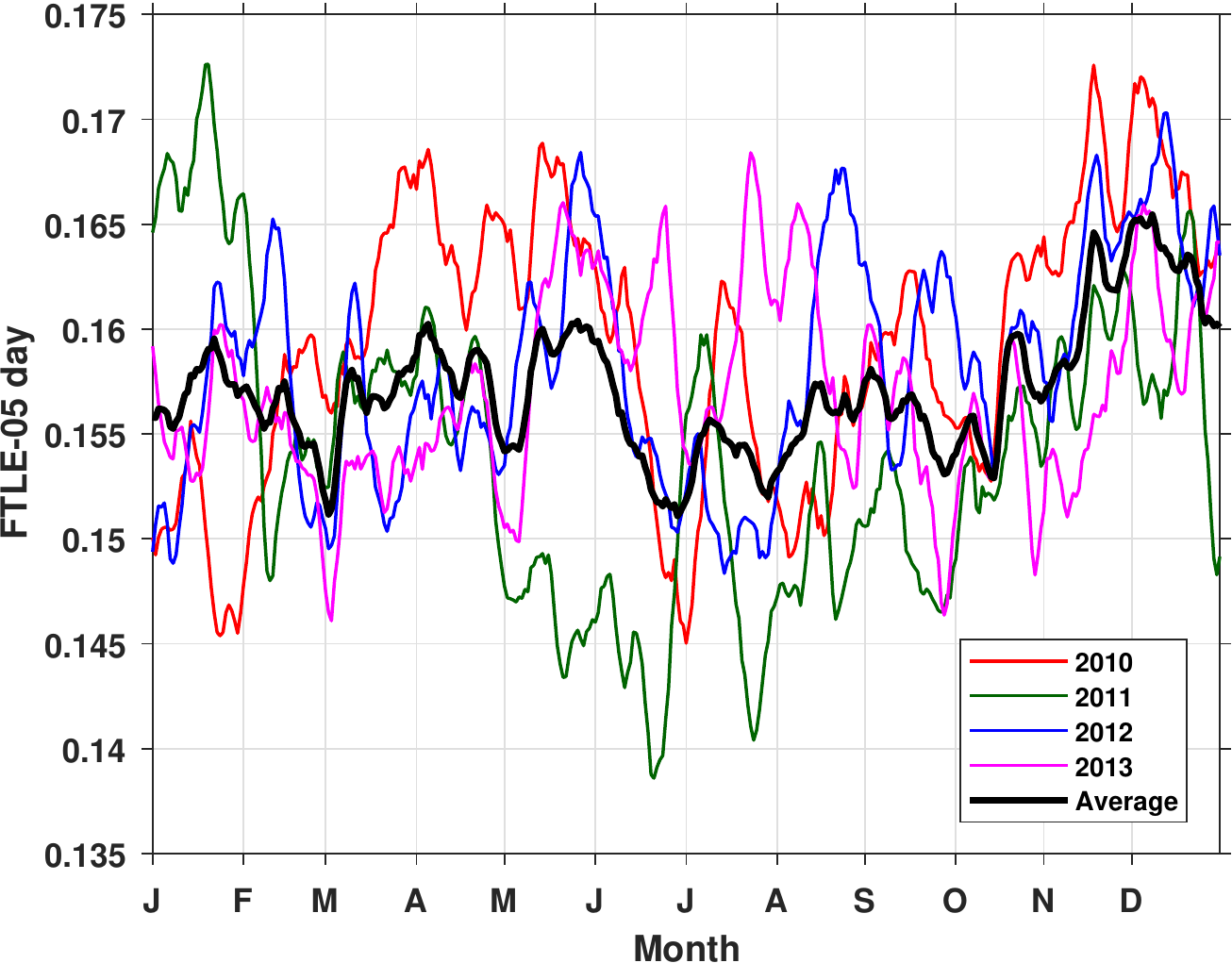}
		}
		\subfigure[]{
		\label{fig 7.3}
		\includegraphics[width=0.4\textwidth]{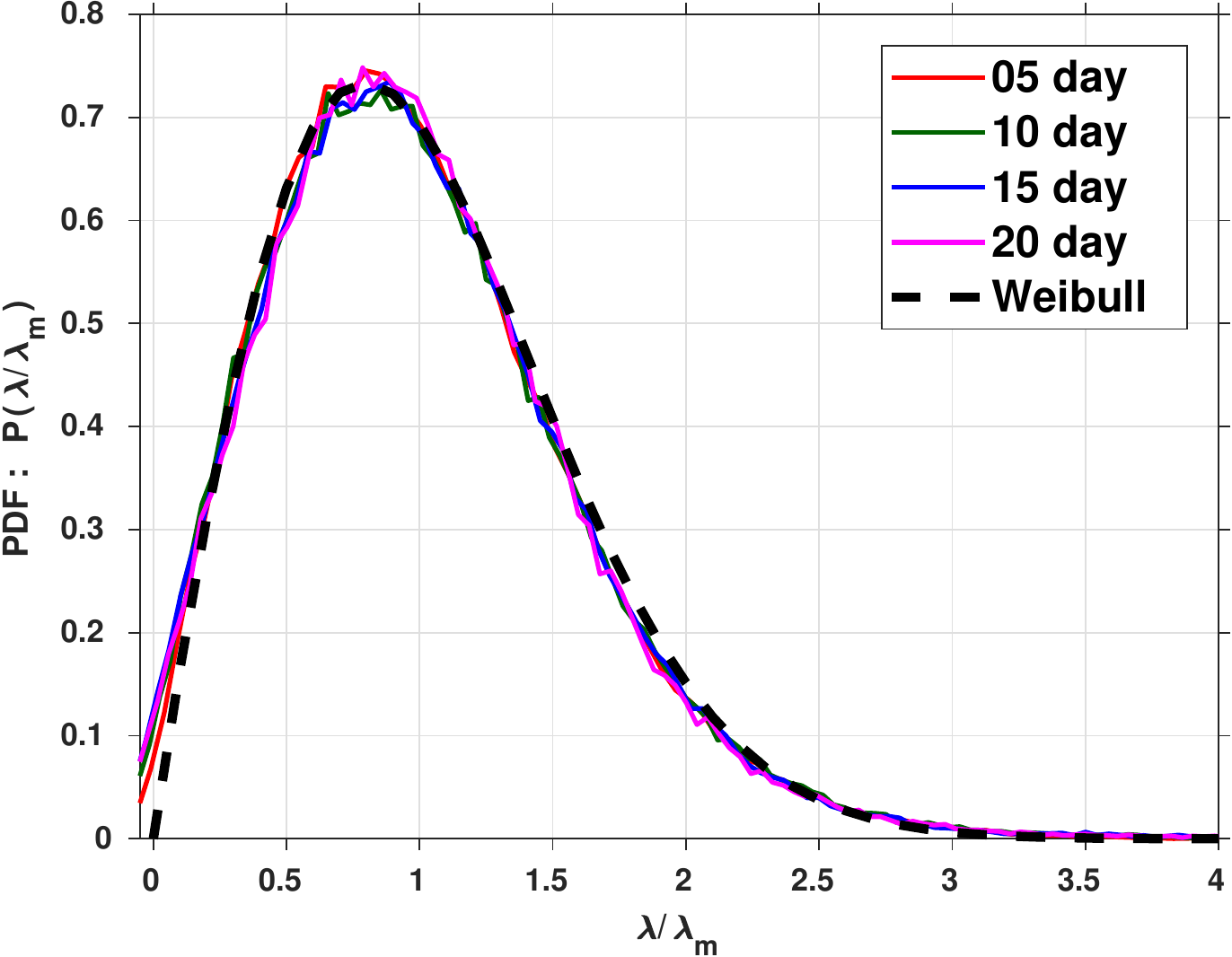}
		}
			
		\caption{Histogram of FTLEs (with different increments) over the Bay; (a) whole year (b) in different seasons. Panel (c) shows the five day mean FTLE time series through the year. Panel (d) shows a fit to the FTLE distribution by a Weibull distribution (normalized by mean $\lambda$ to the FTLE histogram for different $\tau$.} 
		\label{fig7}
\end{figure*}

At a fundamental level, the action of a chaotic flow on the transport or evolution of a passive field requires a knowledge of the probability density function of the FTLEs \citep[see, for example,][]{bf-1999,jai,fh-2004,hv-2004}. The histogram of FTLEs is shown in the first panel of Figure \ref{fig7}. 
This distribution reflects the spatially non-uniform nature of stirring induced by 
the geostrophic currents in the Bay. Indeed, the slow right tail represents regions that experience rapid mixing, while the left tail is indicative of slow rates of stirring. Also, the shape of the distribution changes for different $\tau$. 
Specifically, the histogram becomes taller (mean decreases) with more of a stretched exponential tail for larger $\tau$. In essence, the non-uniformity of mixing is highlighted more starkly for longer time intervals. Even though the regions of strong mixing vary from season to season, as seen in the second panel of Figure \ref{fig7}, there is not much of a seasonal dependence in the FTLE distributions, i.e., the Bay is non-uniformly chaotic throughout the year. This can also be seen in Figure\ref{fig 7.4}, which shows a daily time series of the mean FTLE over the Bay through the year. The average stirring rate is quite uniform through the year, though there is a marginal increase during the post monsoonal months. 
		
For consistency with other parts of the world's oceans  \citep{waugh2008stirring,waugh2012diagnosing}, we note that a Weibull distribution (shown in the fourth panel of Figure \ref{fig7}) accurately fits the FTLE histogram normalized by the mean FTLE. The specific expression plotted in Figure \ref{fig 7.3} is, 
	
\begin{equation}\label{eq:3}
	P(\lambda)=\frac{b}{a}{(\frac{\lambda}{a})}^{b-1}\exp(-\frac{\lambda^b}{a^b}),
\end{equation} 

with $a=1.15$ and $b=1.95$. 
		
\subsection{Relative Dispersion}
		
Another commonly used mixing diagnostic is relative dispersion \citep[RD; see for example,][]{lac1}. This is defined as, 
		
\begin{equation}\label{eq:4}
	\langle R^2(t) \rangle = \frac{1}{N(N-1)}\sum_{i\neq j}^{} {R^2}_{ij}(t)
\end{equation}
		
where, $\langle R^2(t) \rangle$ is the mean relative dispersion of an ensemble of $N$
pairs having the same initial separation with random orientation. 
RD, along with notions such as normal or anomalous diffusion, helps in quantifying the 
homogenization of tracers \citep[see, for example,][for ideal, oceanic and atmospheric applications, respectively]{weeks,lacase,jai-atm}. 
Recently, \cite{waugh2012diagnosing} have compared RD and FTLEs, and highlighted how they explore different aspects of the mixing process. Quite starkly, even in the case when the FTLE distribution collapses to a single point, the RD exhibits a wide spread 
\citep{waugh2012diagnosing}. 
		
\begin{figure*}
	\begin{center}
		\subfigure[]{
		\label{fig 8.1}
		\includegraphics[scale=0.5]{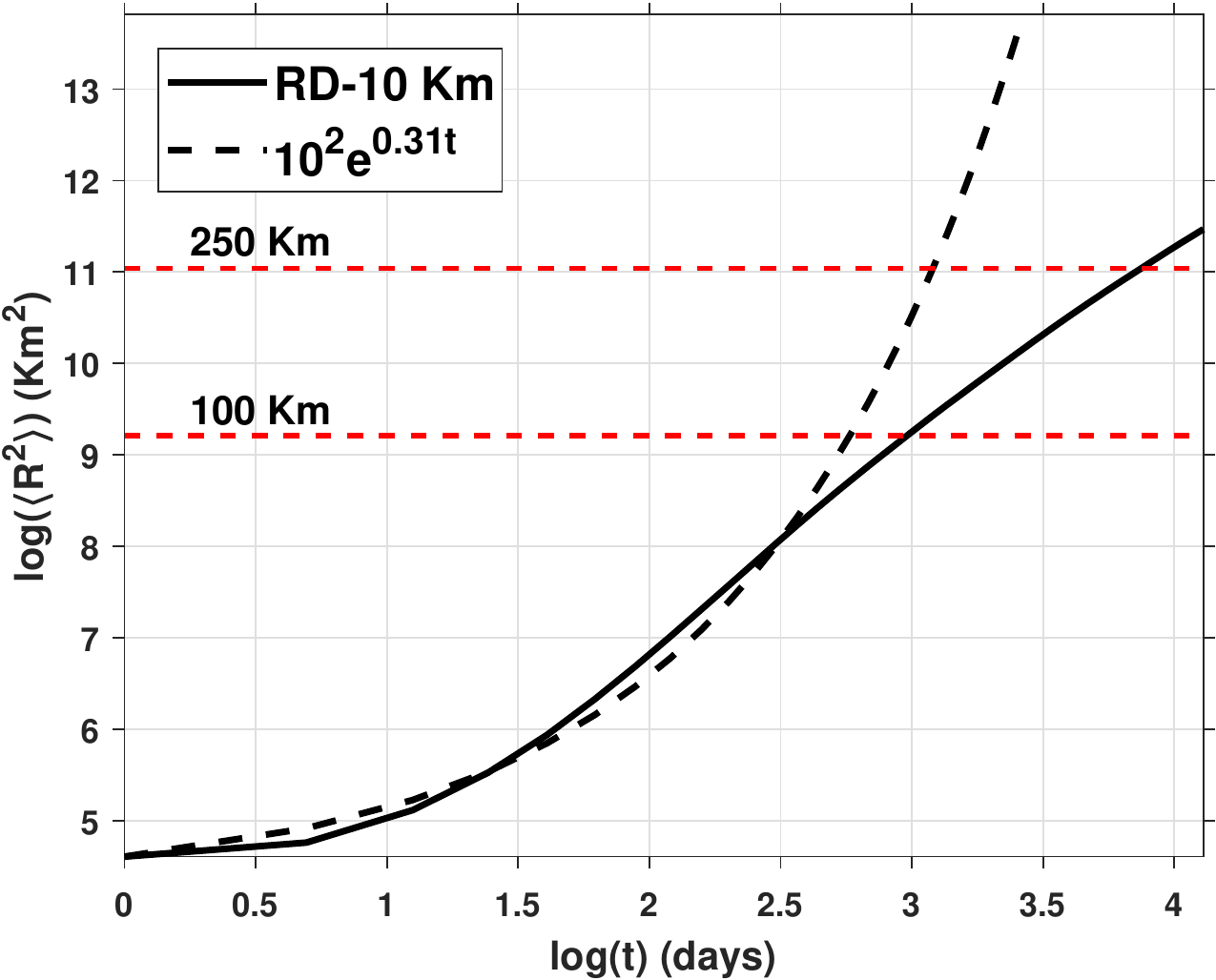}}
		\subfigure[]{
		\label{fig 8.2}
		\includegraphics[scale=0.5]{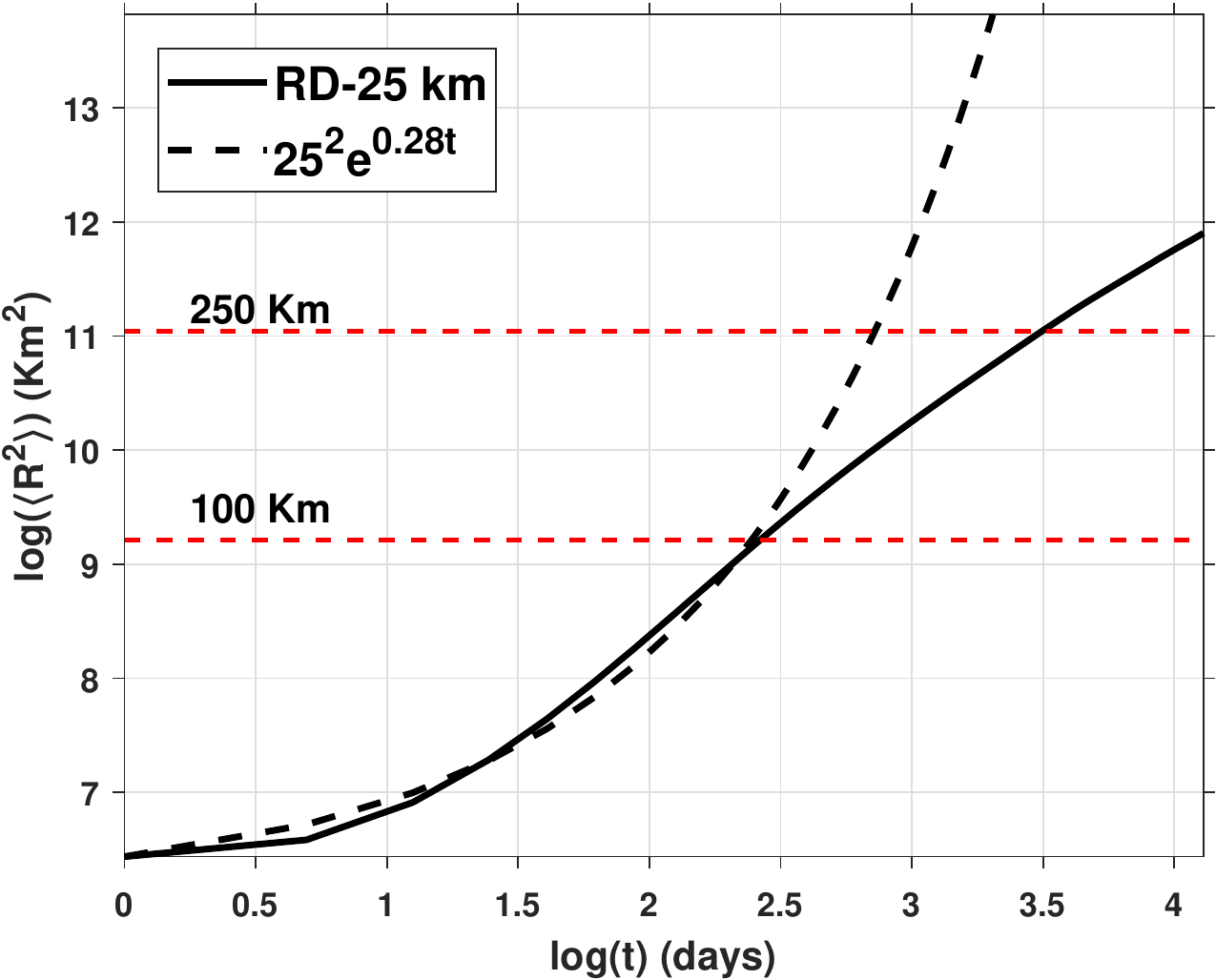}}\\
		\subfigure[]{
		\label{fig 8.3}
		\includegraphics[scale=0.5]{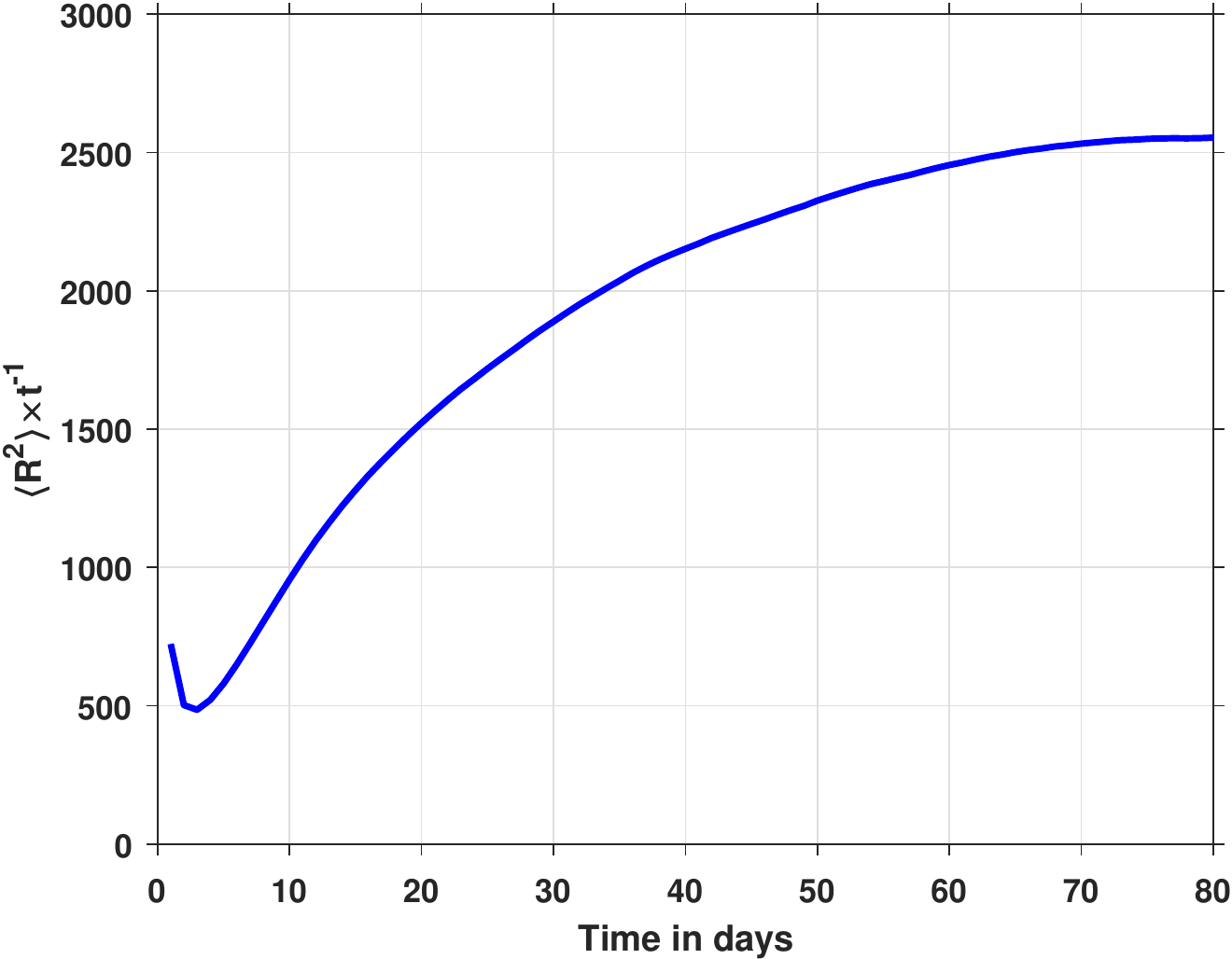}} \\
		\subfigure[]{
		\label{fig 8.4}
		\includegraphics[scale=0.5]{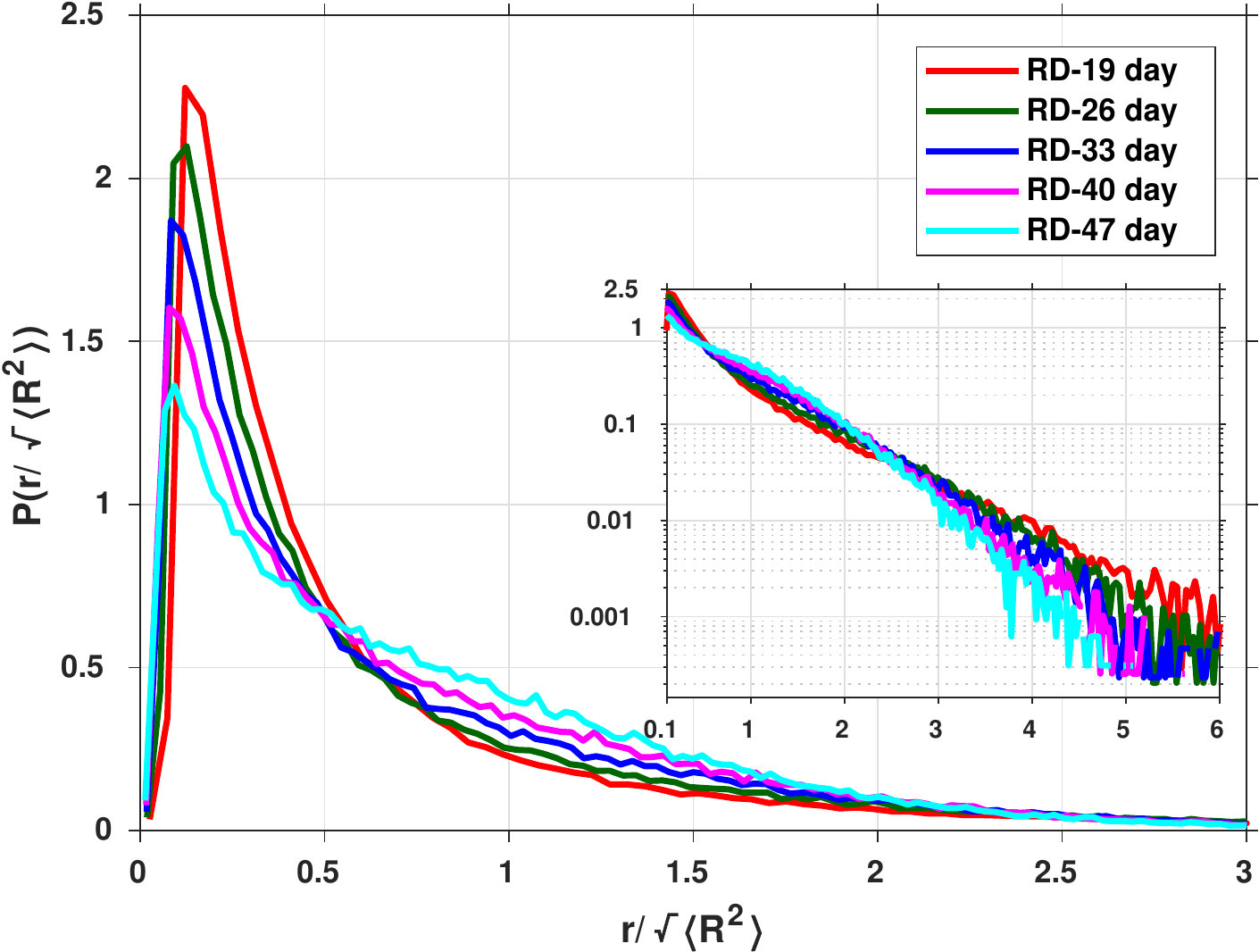}}
		\subfigure[]{
		\label{fig 8.5}
		\includegraphics[scale=0.5]{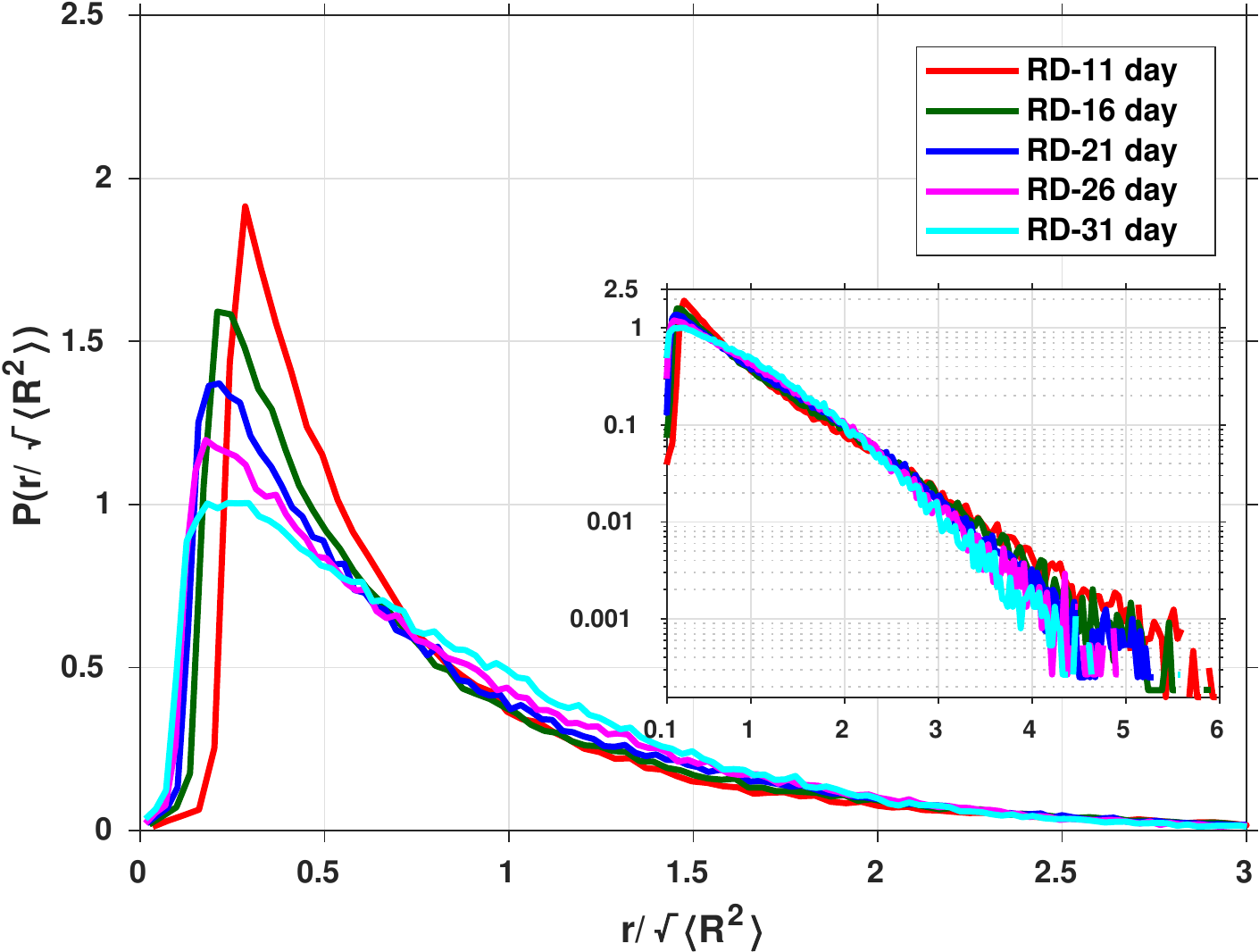}}
		\caption{Panels (a) and (b) show the RD with time for initial separations of 10 and 25 $km$, respectively. Panel (c) shows the compensated RD (by $t^{-1}$) as a function of time. Panels (d) and (e) show histograms of the square root of RD (denoted by $r$, normalized by its rms value) at different days when the mean RD is between 100 and 250 $km$.}
	\end{center}
		\label{fig8}
			
\end{figure*}

In practice, we advected 1000 pairs of parcels at a given grid location for 90 $days$ starting on the first day of each month in all the four years. Two suites of experiments were conducted with pairs in a circle that are initially 10 and 25 $km$ apart, respectively. The evolution of RD with these two initial separations is shown in Figure \ref{fig 8.1} and \ref{fig 8.2}. At small scales, the advecting flow is a smoothly interpolated geostrophic flow, hence the observed exponential separation is not surprising. As discussed earlier, at small scales (below approximately 100 $km$, depending on the region in consideration), the surface ocean flow appears to have a relatively shallow kinetic energy spectrum \citep{callies}, and a significant divergent component \citep{capet2008mesoscale,BCF,qiuetal}. 
In fact, pair separation on the order of a few $km$, estimated using drifter data from the Gulf of Mexico, appears to conform with the classic Richardson $\langle R^2 \rangle \sim t^3$ prescription \citep{Poje-etal}. Indeed, it would be interesting to ascertain the behavior of RD in the 10-100 $km$ range via high resolution surface ocean models.
		
At scales between 100 and 250 $km$, we observe that the RD is reasonably well approximated by a power-law growth in time. While we anticipate exponential separation in an enstrophy cascading regime, it should be remembered that this is under the assumption of an inertial range with a constant enstrophy flux \citep{lin}.
Computation of the enstrophy flux using altimeter data does show a dominant forward enstrophy regime at these scales, but the enstrophy flux is not constant, and is actually scale dependent \citep{khatri}. Therefore, the power-law growth of RD is not 
inconsistent with a forward enstrophy transfer regime. Finally, at scales larger than 250 $km$, the growth of RD slows down and tends to $\langle R^2 \rangle \sim t$, indicative of eddy diffusive behavior. This is seen in the emergent flat portion of a compensated RD plot (for the 25 $km$ initial separation case) beyond approximately 70 $days$ of evolution as shown in Figure \ref{fig 8.3}.
		
\begin{figure*}
	\begin{center}
	\includegraphics[scale=0.6]{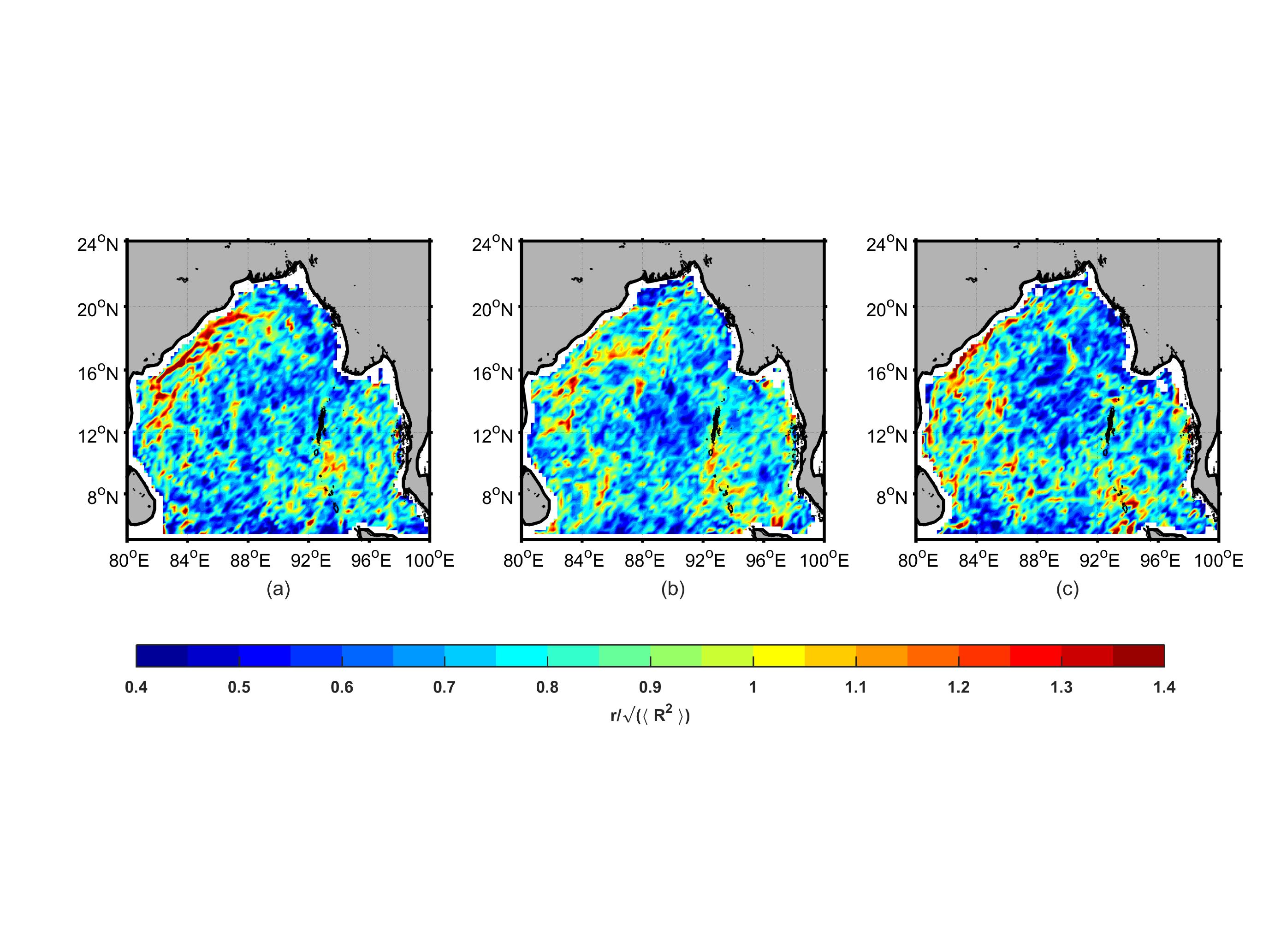}
	\caption{Panels (a), (b) and (c) show maps of the square root RD (normalized by its rms value) at 15 $days$ from an initial cluster of 25 $km$ separation in the premonsoon (FMA), monsoon (JJAS) and postmonsoon (OND) periods, respectively.}
	\end{center}
	\label{fig9}
\end{figure*}

As with the FTLEs, the distribution of RD at a given time is of importance in assessing the non-uniformity of mixing. The distribution of RD (actually, the square root of RD, normalized by its rms value), at various times, when the mean square root RD lies between 100 and 250 $km$ is shown in Figure \ref{fig 8.4} and \ref{fig 8.5} for 10 and 25 $km$ initial separations, respectively. 
Quite clearly, the distributions are not Gaussian, rather they show signs of a log-normal behavior. Spatial maps of the mean RD are presented in Figure \ref{fig9}. In the premonsoon period (FMA), the western coastline supports rapid pair dispersal, while the southern Bay shows subdued RD values. In the monsoon season (JJAS), high RD shifts slightly off the western coast, and in addition, rapid dispersion picks up in the parts of the southern and eastern Bay. Finally, in the postmonsoon period (OND), along with the west coast, much of the southern and eastern Bay lights up with high RD values. Interestingly, all through the year, the central Bay is characterized by low RD values, and this also points to a local or eddy scale mixing, rather than basin wide homogenization.
		
\subsection{Finite Size Lyapunov Exponent}
		
For a flow with multiple scales, such as that in the BoB, the Finite Size Lyapunov Exponent (FSLE) is a convenient measure that quantifies the chaotic nature of the flow 
\citep{artale1997dispersion,hernandez2011reliable}. In particular, the FLSE allows one to estimate the scale up to which mixing is chaotic, and beyond which, a diffusive framework is appropriate. From a practical viewpoint, this also allows an estimation of a scale dependent diffusion coefficient \citep{Lacorata}. 
		
Following \cite{garcia2007dispersion}, a set of $M$ tracers with some initial distribution and standard deviation $\sigma$ are followed in time as they are transported by the velocity field. The parameter $\sigma(t)$ is defined as,
		
\begin{equation}\label{eq:5}
	\sigma(t) = {\langle{|x_{i}(t) - \langle x_{i}(t)\rangle|}^2 \rangle}^{1/2}, \textrm{where},
\end{equation}
		
\begin{equation}\label{eq:6}
	\langle x_{i}(t)\rangle \equiv \langle \{x_{i}(t) : i = 1, 2, ... M\} \rangle= \frac{1}{M}\sum_{i=1}^{M}x_{i}(t).
\end{equation}
		
We define the initial size of the cluster ${\sigma}_{0}$ according to \ref{eq:5} and \ref{eq:6}, and measure the time $T$(${\sigma}_{0}$,r) it takes for the growth from ${\sigma}_{0}$ to ${\sigma}_{1}$ = r $\sigma_0$ , $T$($\sigma_1, r$) the time 
it takes for the growth from $\sigma_1$ to $\sigma_2$ = r$\sigma_1$, and so on up to the largest scale under consideration (the sub-basin scale). A set of $N$ experiments is then performed with different initial conditions for the cluster of particles $N \gg$ 1, and we calculate the mean time $\tau$ that a cluster with size $\sigma_j$ takes to grow by a factor $r$ and the $\langle \cdot \rangle$ operation is the average over the tracer ensemble. This reads, 
		
\begin{equation}\label{eq:7}
	\tau(\sigma_j,r) =\langle T(\sigma_j,r)\rangle = \frac{1}{N}\sum_{j=1}^{N}T(\sigma_j,r).
\end{equation}\
		
The FSLE parameter as a function of the scale is then obtained by the following expression, 
	
\begin{equation}\label{eq:8}
	\lambda(x, {\sigma}_j ) = \frac{\log r}{ \tau(\sigma_j,r)},
\end{equation}
		
which is not sensitive to variations in $r$ when it is close to 1+. From the FSLE, we define the Finite Size Diffusion Coefficient (FSDC, $D$) as,
		
\begin{equation}\label{eq:9}
	D(\sigma) = {\sigma}^2 \lambda(\sigma).
\end{equation}
		
As the cluster size grows, if it is driven by chaos at very small scales ($ \ll l_u$) and by (eddy) diffusion at very large scales ($\gg l_u$), then $\lambda(\sigma)$ has the following asymptotic behavior, 
	
\begin{equation}\label{eq:10}
	\lambda(\sigma)= 
	\begin{cases}
	{\lambda}_{max}, & \text{if}\ \sigma \ll l_u, \\
	\frac{D}{{\sigma}^2}, & \text{if}\ \sigma \gg l_u. \\
	\end{cases}
\end{equation}
		
Though, for real world data, as demonstrated in \cite{Corr} in an oceanic setting, there usually exist other intermediate regimes for $\lambda(\sigma)$. 
		
Practically, for calculating the FSLE, a cluster with 10000 parcels was released at each grid location, and these were advected in time for approximately 60 $days$. In fact, we have chosen a cluster whose initial standard deviation is 25 $km$ and followed the parcels till the standard deviation increases by a factor of 1.2. The size of the initial cluster is varied from 25 $km$ to approximately 300 $km$.
This experiment was repeated for every year, and an average was taken over all years.
		
\begin{figure*}
	\begin{center}
		\subfigure[]{
			\label{fig 10.1}
			\includegraphics[scale=0.6]{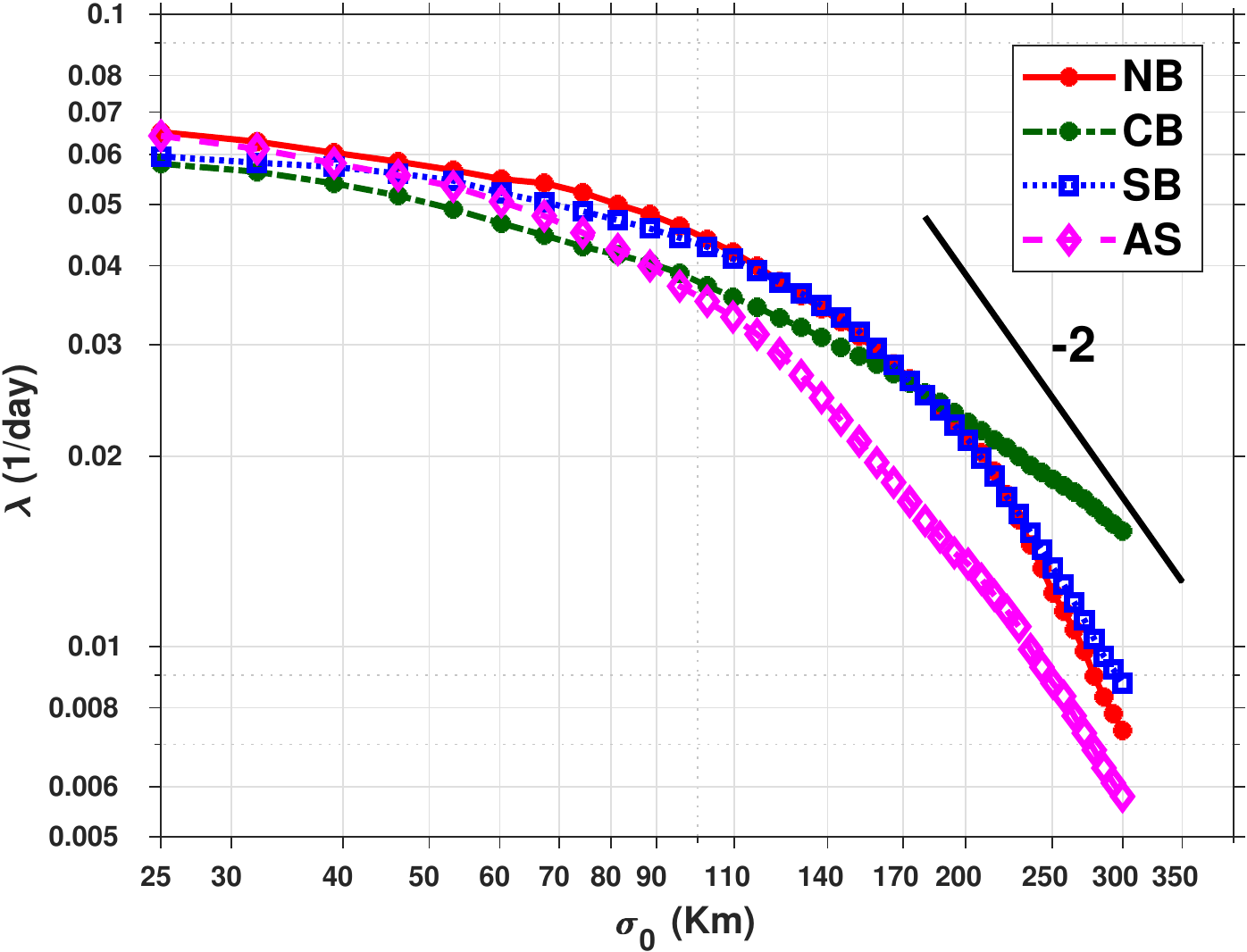}}
		\subfigure[]{
			\label{fig 10.2}
			\includegraphics[scale=0.6]{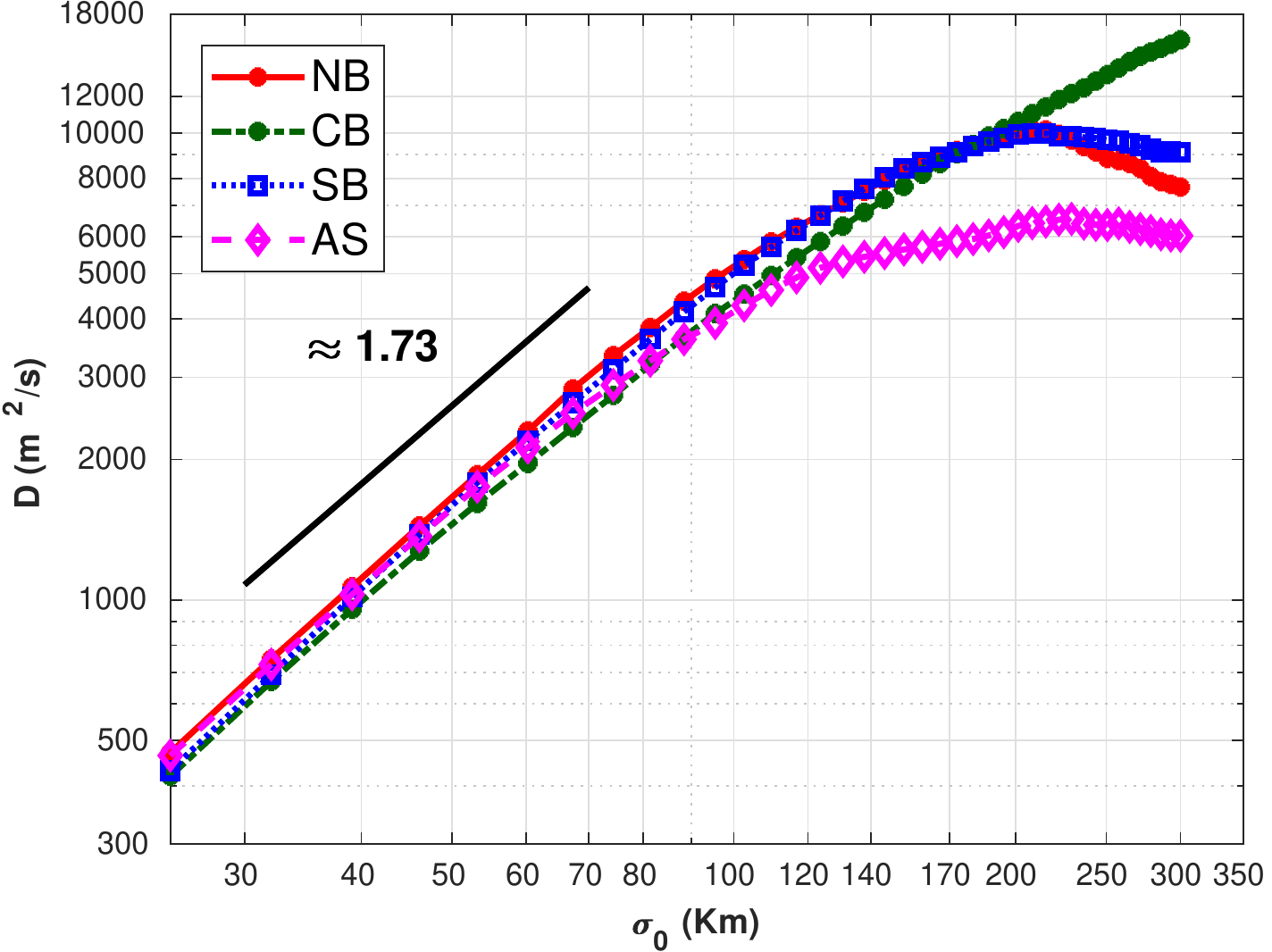}}
			\caption{Panels (a) and (b) show FSLE and eddy diffusivity as a function of initial cloud size with expansion factor $r=1.2$ for the Northern Bay, Central Bay, Southern Bay and Andaman Sea.}
	\end{center}
			\label{fig10}
\end{figure*}

The behavior of the FSLE with cluster size can be seen in Figure \ref{fig 10.1}. As with RD, the smoothly interpolated flow results in an exponential separation, i.e., a relatively flat FSLE, for scales below approximately 100 $km$. For the NB and SB an eddy-diffusive regime is observed for scales greater than 200-250 $km$ (via a $-2$ scaling of the FSLE in Figure \ref{fig 10.1}). At intermediate scales, i.e., 100 to 250 $km$, the FSLE transitions between these two end regimes. The behavior of the Andaman Sea region is somewhat different with the diffusive regime emerging earlier between 150 and 200 $km$, itself. This overall picture is consistent with 
the calculated ``eddy length scale" and the local chaotic mixing seen in Figure \ref{fig9}. Interestingly, the CB is quite distinct in that the FSLE never really transitions to an eddy-diffusive regime. It is possible that, especially in the intermediate regime, finer data may yield different power-laws as seen by \cite{Corr} in other parts of the world's oceans. An estimation of a scale dependent diffusion co-efficient, the FSDC, is presented in Figure \ref{fig 10.2}. It is interesting to note that the FSDC scales as a power-law with cluster size (exponent of 1.73), this is potentially a useful reference which provides a resolution-diffusivity relation for use in models. At large scales, for the NB and SB, the eddy diffusivity is approximately $10^4$ $m^2/s$. In the Andaman Sea region, the value is lower and is near $6 \times\! 10^3$ $m^2/s$. These numbers are higher (by a factor of 2) than the minimum Osborn-Cox eddy diffusivity estimates as determined by \cite{AbM} (see in particular the Bay of Bengal in their global Figure 5), but are of the same order as diffusivities estimated via FSLEs in other parts of the world's oceans \citep{Corr}.
		
\begin{figure*}
	\begin{center}
		\includegraphics[scale=0.6]{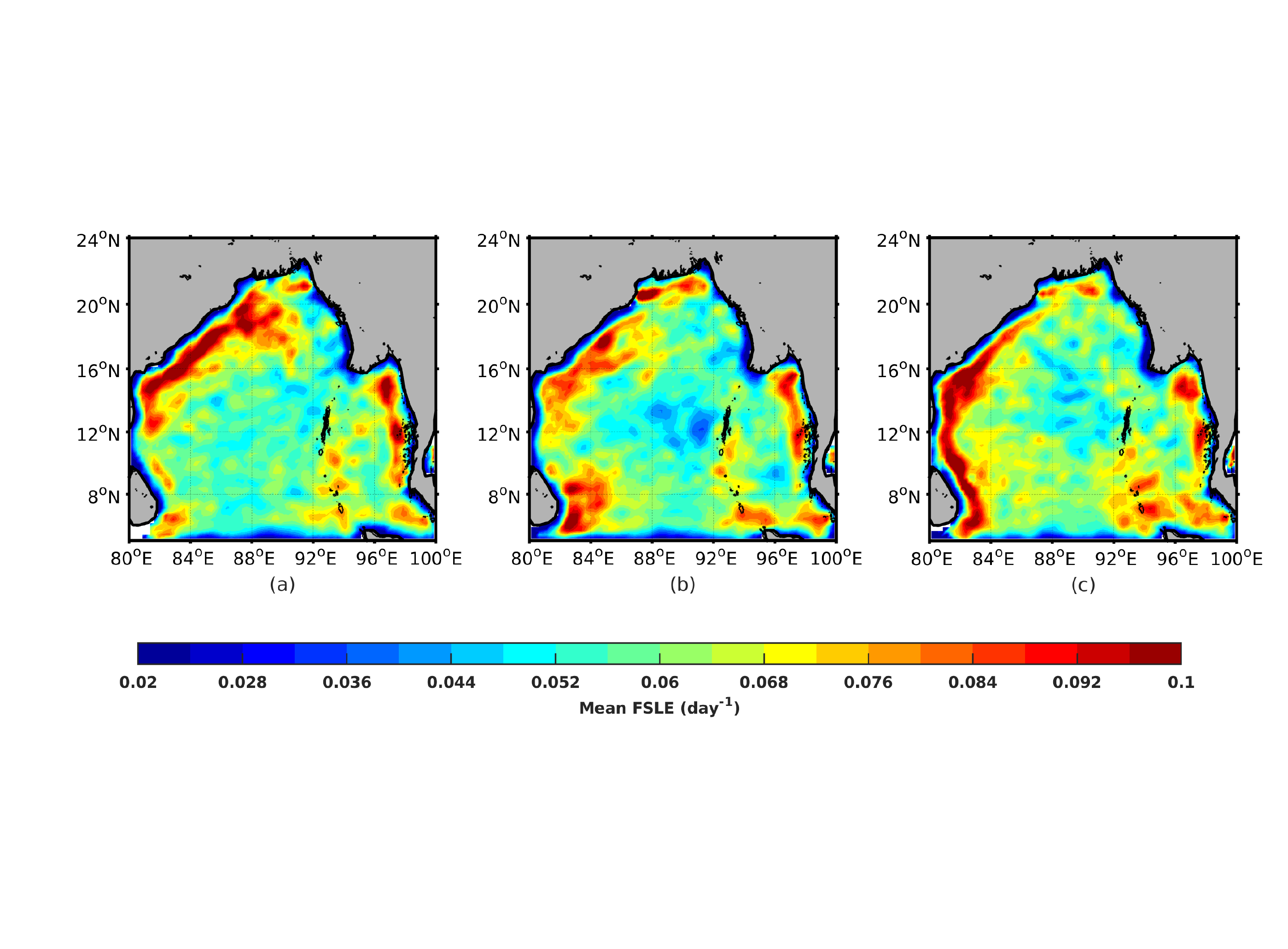}
		\caption{Panels (a), (b), (c)  show the mean FSLE for FMA, JJAS, OND seasons respectively. This is an average over all four years of altimetry data.}
		\label{fig11}
	\end{center}
\end{figure*}

The seasonal mean picture of FSLEs (within the chaotic regime) is shown in Figure \ref{fig11}. As with the seasonal picture from the FTLEs, high values of the FSLE show equatorward movement from FMA, through JJAS into OND. Note that the mouth of the river Ganga-Brahmaputra basin and Irrawaddy catchment area show high FSLE in JJAS and OND season. Signs of the Sri Lankan and BoB domes can also be seen in the FSLE maps in JJAS and OND, respectively. Also, low FSLE values can be seen in the CB region throughout the year which are consistent with a partition that inhibits basin scale mixing. 	

\section{Analysis of a single fresh water mixing event}
		
As mentioned earlier, a significant amount of surface fresh water enters the Bay in the postmonsoon season from the Ganga-Brahmaputra (GB)
catchment \citep{papa2012ganga,chait}, and plays an important role in its surface salinity budget \citep{raghu}. This inflow from GB progresses southward and is largely confined to the western side of the Bay \citep[referred to as a river in the sea,][]{chait}. The fate of this fresh water, including its advection, and 
homogenization with a more salty environment has been analyzed in detail \citep{Akil}. Indeed, it appears that the fresh water maintains its identity for an extended period and finally extended tongues become saltier in the southern Bay \citep{Bens,Akil}. Lagrangian salinity change maps from data from August to October of 2013 also show an increase in the saltiness with southward advection \citep{Amala}. Though it should be kept in mind that some fresh water retains its identify for a prolonged period and its transport to remote regions in the Indian Ocean has been documented \citep{Deb1}. 
		
\begin{figure*}
	\begin{center}
		\subfigure[]{
			\label{fig 12.1}
			\includegraphics[scale=0.6]{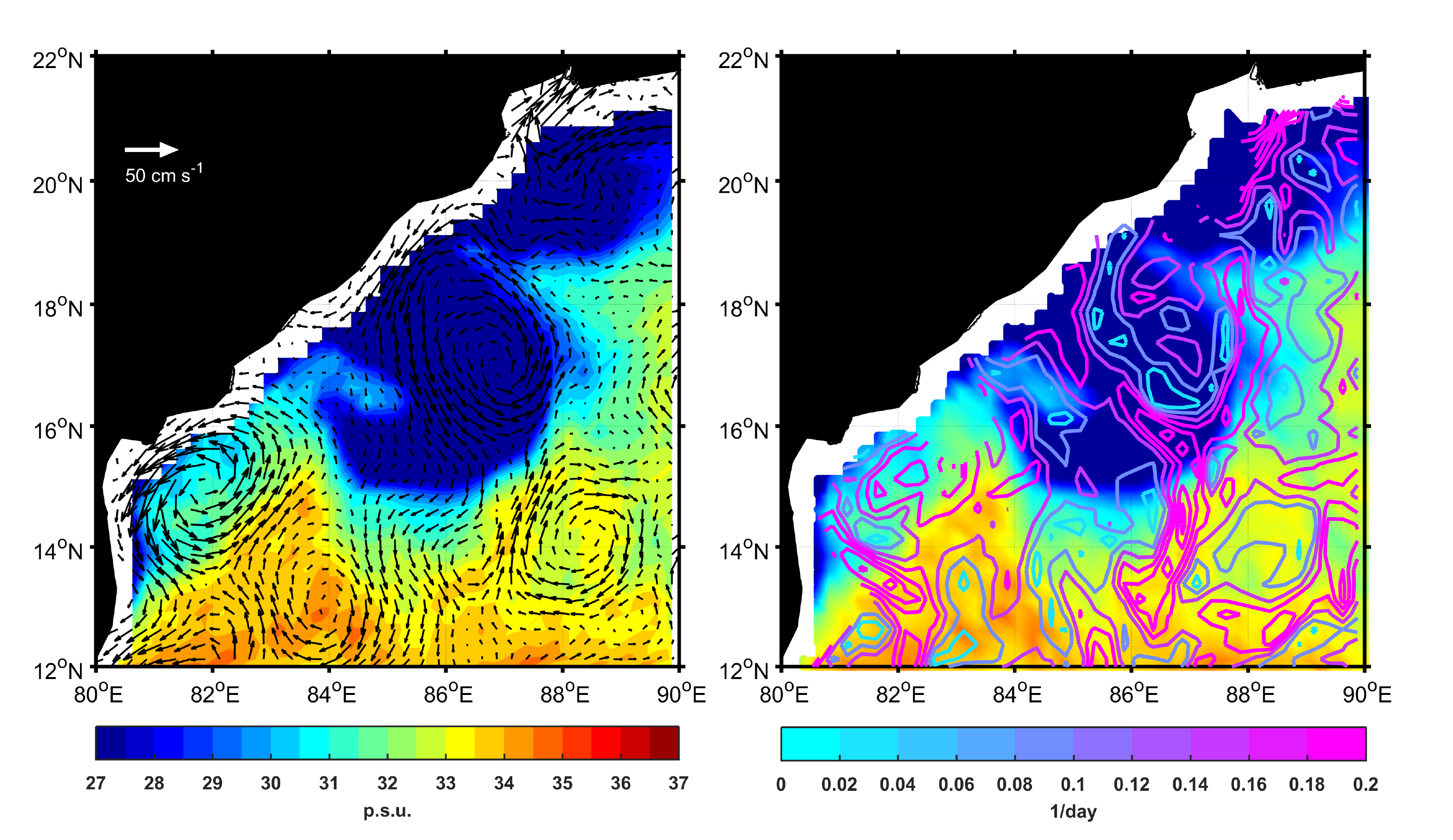}}
		\subfigure[]{
			\label{fig 12.2}
			\includegraphics[scale=0.6]{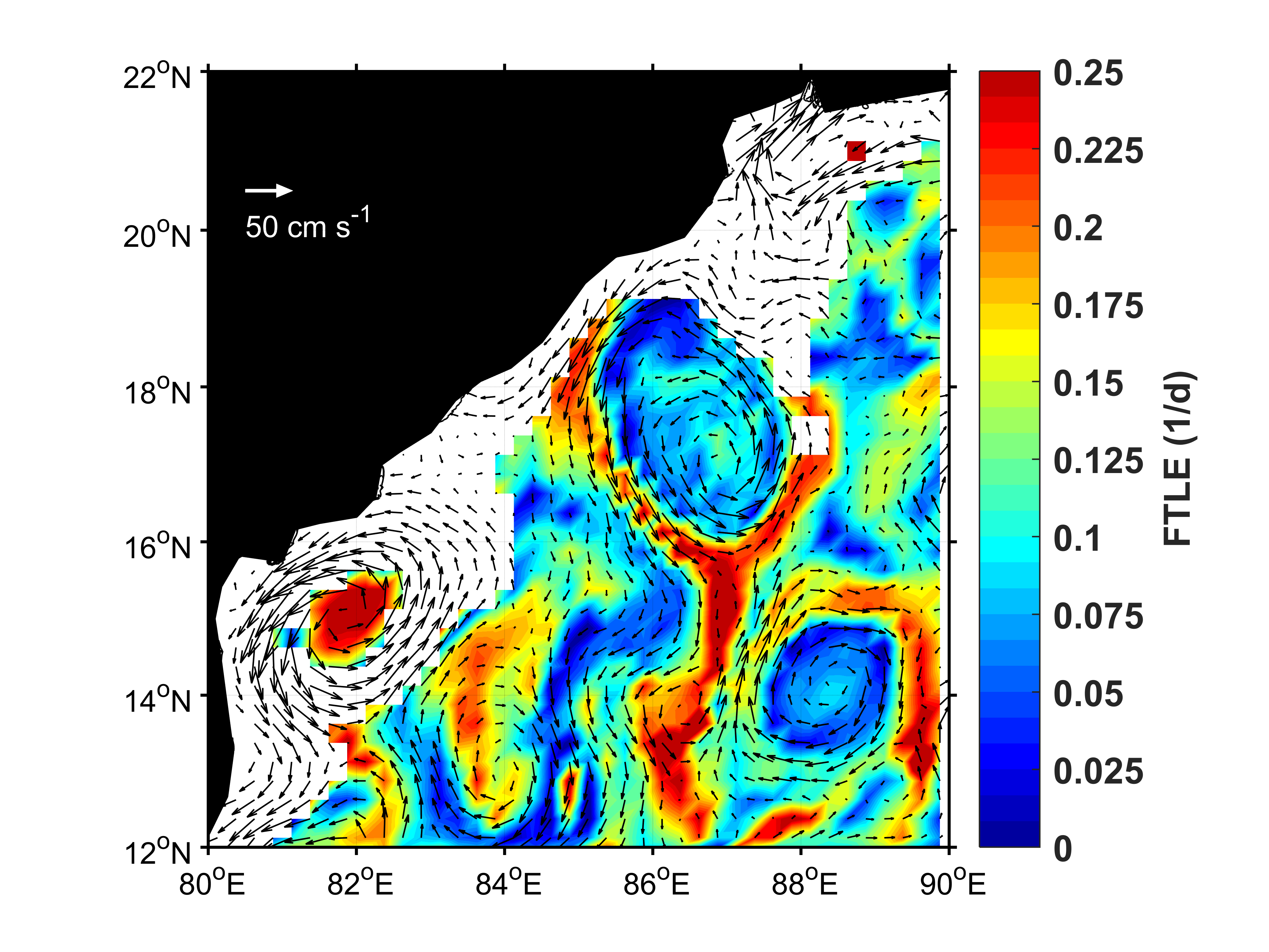}
			\includegraphics[scale=0.6]{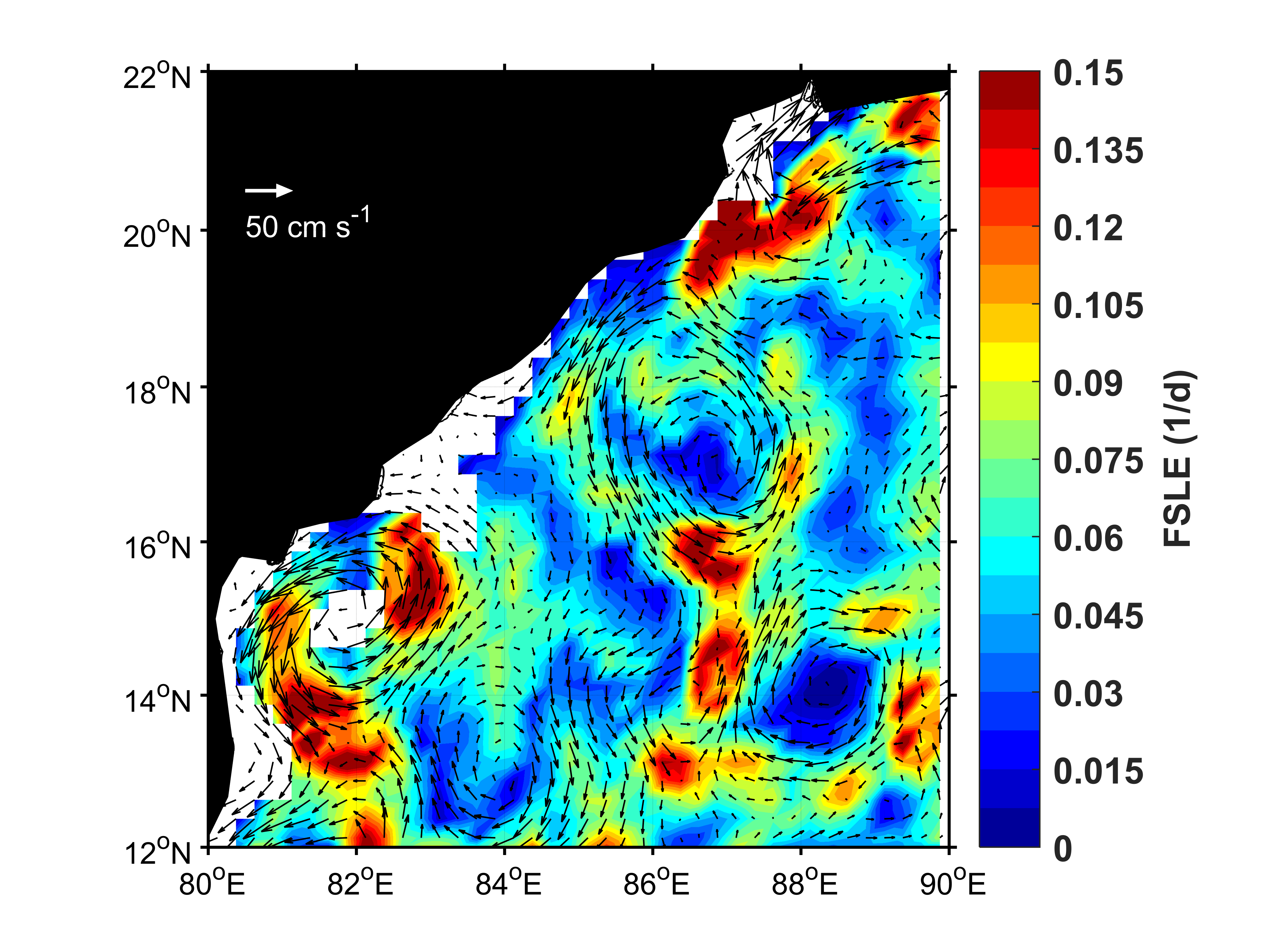}}
			\caption{Panel (a) shows salinity map with velocity quivers (left) and FTLE-05 day contours overlaid onto passive scalar field (right) on 25/10/2015. Panel (b) shows the spatial map of FTLE-15 day (left) and FSLE (right) on 25/10/2015.}
	\end{center}
	\label{fig12}
	\end{figure*}

Taking advantage of satellite data from 2015, stirring of the salinity field by eddies in the postmonsoon season has been recently explored by \cite{Chait1}.
Here, we highlight role of eddies off the eastern coast on India in preserving the identity of fresh water. In particular, we showcase a particular eddy that formed around the end of October and lasted to the end of November in 2015. The salinity field along with quivers of the geostrophic flow on October 25 are shown in 
the first panel of Figure \ref{fig 12.1}. The eddy is well formed and spans the approximate region 16-19$^\circ$N to 85-88$^\circ$E, with a slight 		northwest-southeast tilt. 
The second panel of Figure \ref{fig 12.1} shows a passive tracer initialized to have the same value as the salinity field, along with contours of 5 day FTLEs. Note that a fair amount of fresh water (blue) is contained within the eddy. The two panels of Figure \ref{fig 12.2} show maps of FTLEs (15 $day$) and FSLEs ($r=1.2$) in this part of the Bay. Quite clearly, the region inside the eddy is characterized by relatively low FTLEs/FSLEs while it is surrounded by a high FTLE/FSLE ring. Also, note that inside the eddy, the interior rim has lowest FTLEs while the central region has relatively higher stirring rates. Thus, much like the scenario described by \cite{RY-1983}, we expect a tracer to be well mixed within an eddy, but not communicate significantly (or, possibly on a slower timescale) with the ``external" Bay. 
		
\begin{figure*}
	\begin{center}
		\subfigure[]{
			\label{fig 13.1}
			\includegraphics[scale=0.6]{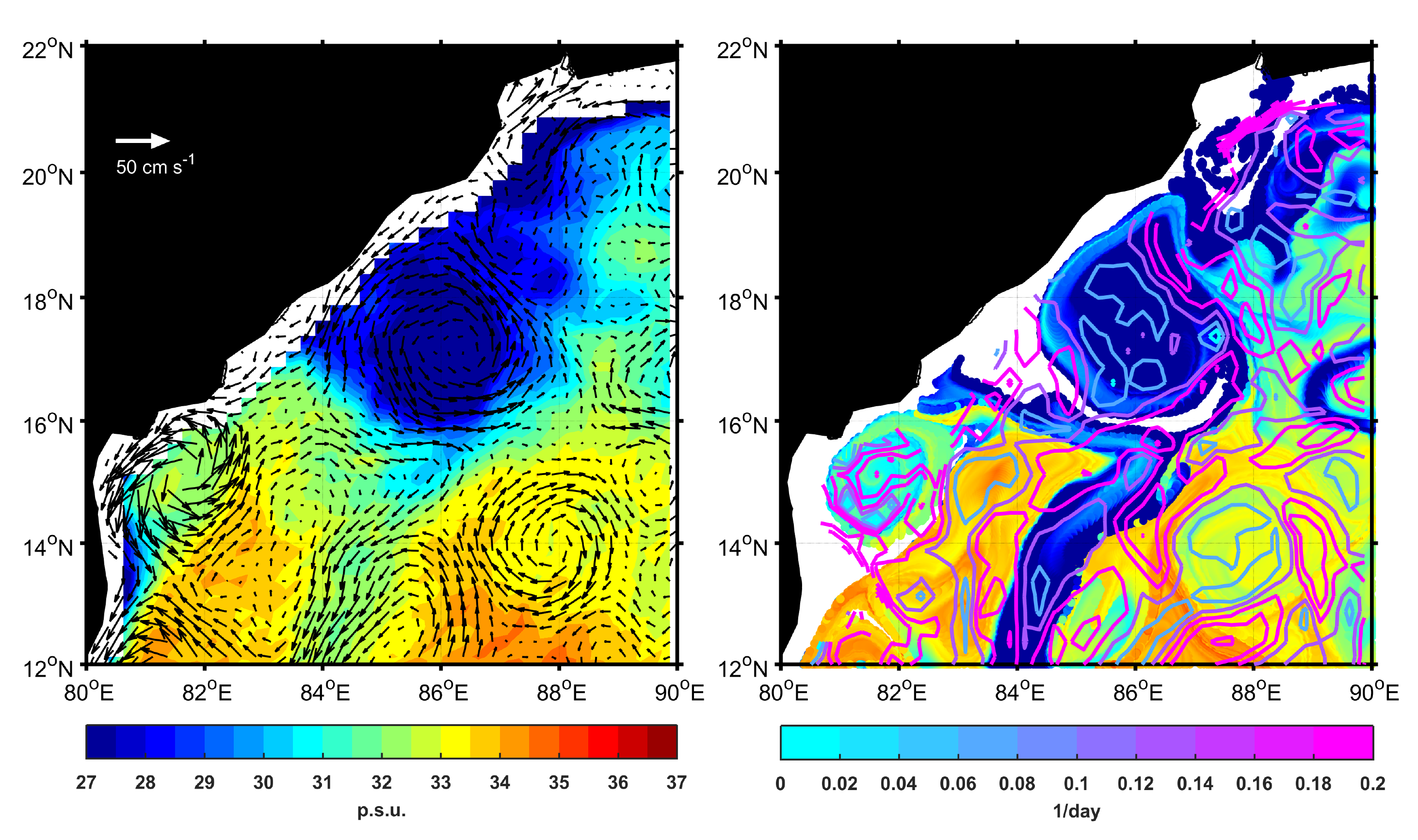}}
	  	\subfigure[]{
			\label{fig 13.2}
			\includegraphics[scale=0.6]{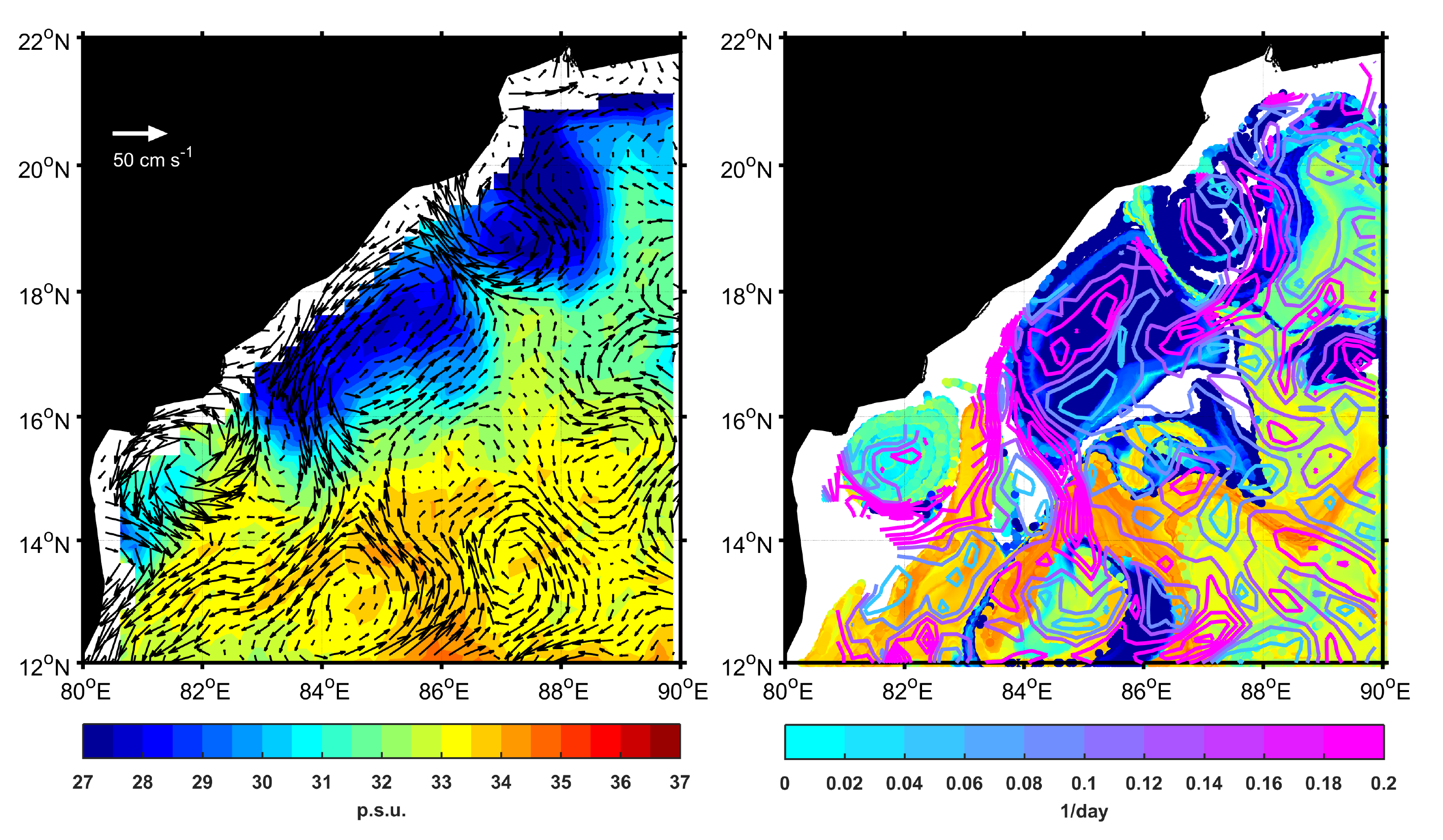}}
	  		\caption{Panels (a) and (b) shows salinity map with velocity quivers (left) and FTLE-05 day contours overlaid onto passive scalar field (right) on 08/11/2015 and 24/11/2015, respectively.}
	\end{center}
		\label{fig13}
\end{figure*}

Fifteen days later, as seen in the first panel of Figure \ref{fig 13.1}, low salinity water is retained within the eddy. The passive tracer, advected via the geostrophic flow in these fifteen days is shown in the second panel of Figure \ref{fig 13.1}. While the bulk of ``fresh" passive tracer is in the eddy, long filaments that initially comprised of material localized on the outer side of the eddy, extend southwards up to 12$^\circ$N and westward to the coast. These narrow filaments are not seen in the actual salinity field, and have most likely been homogenized via the more salty environment --- possibly due to strong vertical mixing \citep{Bens,Akil}. Finally, another fifteen days later, on November 24 (Figure \ref{fig 13.2}), the passive and salinity fields compare favorably, even though the eddy itself has deformed considerably. In fact, it is now squeezed
along the coast into a northeast-southwest orientation, and fresh water begins to be strained out of its southwest boundary. Remnants of ejected filaments from the edges of the eddy that are seen in the passive field are absent in the salinity data, presumably also having being been homogenized to more salty levels.
		
Thus, in this period of one month, fresh water trapped inside the eddy is shielded by means of kinematic transport barriers that are clearly delineated by the FTLE and FSLE maps. In addition, this eddy trapped fresh water stays in place rather than being rapidly advected in a southward direction. On comparison with the passive tracer, we note that ejected passive filaments are not seen in the fresh water signature, thus highlighting the role of the kinematic barriers in the preservation of fresh water on this monthly timescale.
		
\section{Summary and Discussion}
		
Using surface geostrophic currents derived from altimetry data, we studied mixing along the surface of the Bay of Bengal. In particular, 
our primary focus was on mixing that takes place on intraseasonal scales. To begin with, we examined the flow by means of 
Hovm{\"o}ller plots and wavenumber-frequency diagrams. It was seen that the geostrophic currents in the Bay are 
dominated by westward progressing disturbances that have temporal scales between 50 and 100 $days$. In fact, the power in these systems aligned well
with the theoretical dispersion curves for linear baroclinic Rossby waves. Interestingly, some of these have length scales that are smaller than the 
local deformation scale, and show an eastward group velocity which was noted in the Hovm{\"o}ller plots. 
Temporal and spatial power spectra were seen to follow approximate power-laws ($-3$ scaling, from 100-250 $km$ and 10-30 $days$, respectively) and suggested 
an uninterrupted distribution of power across length and subseasonal time scales. 
		
The advection of latitudinal and longitudinal bands by the multiscale geostrophic flow immediately hinted at the presence of chaotic mixing. In particular, the repeated folding and filamentation of stripes brought forth a complicated geometry to the mixing process, and by the end of approximately six weeks, it was difficult to distinguish between the two initial conditions. In addition, mixing was not basin wide but was seen to be restricted to the scale of eddies. 
		
A more quantitative measure of mixing was provided by the FTLEs. Maps of the FTLEs suggested an equatorward movement of regions of enhanced mixing from pre-monsoonal to post-monsoonal periods. In each season, the central Bay had low FTLE values, suggestive of the presence of kinematic barriers that were consistent with the eddy scale of mixing noted above. Specific seasonal features, such as the appearance of the Sri Lankan dome in the monsoon, were captured by high
FTLE pockets. Also, overall, the spatial maps of FTLEs were in tune with those of eddy kinetic energy. Variations in FTLEs are known to be important in determining the outcome of advection-diffusion on passive fields; here, the non-uniform 	nature of surface mixing in the Bay was manifest in the histograms of FTLEs that had long tails and their shape, and as in other parts of the world's oceans, was captured well by a Weibull distribution. In terms of a domain average, the FTLE for a week's increment was approximately 0.1 $day^{-1}$, while the spread captured by the histogram ranged up to $0.5$ $day^{-1}$. In addition, with longer time increments, the distribution of FTLEs became taller (and smaller mean), but with progressively more stretched exponential like tails. Thus, the non-uniformity of mixing was further highlighted at longer time intervals. 
		
The relative dispersion (RD) of parcels provided a complimentary view to the FTLEs. Below 100 $km$, the smoothly interpolated nature of the data results in pair separation that was exponential in time. From 100 to 250 $km$, i.e., the RD followed a power-law in time, which is consistent a forward enstrophy transfer regime, but with a variable enstrophy flux. At larger scales, the pair separation took on an eddy-diffusive growth, i.e., $\langle R^2 \rangle \sim t$.
Consistent with the FTLEs, seasonal mean maps showed a southward progression of high RD, and a suppressed dispersion in the central Bay throughout the year.
		
FSLEs were then estimated for the Bay, these provide a quantitative measure of the scale up to which tracers experience chaotic mixing. Averaging the growth of clusters in different months, and across all four years, the FSLE followed theoretical expectations. In particular, the FTLE was relatively constant (up to 100 $km$), transitions (from 100 to 250 $km$) and then enters an eddy-diffusive regime (above 250 $km$). The Andaman Sea was seen to enter an eddy-diffusive regime at relatively smaller scales (150 $km$) while the central showed no signs of this transition. Overall, in agreement with FTLEs and latitudinal (longitudinal) stripe experiments, chaotic mixing takes place within eddies. Beyond the eddy scale, the $-2$ power-law suggested an eddy diffusive behavior (except in the central Bay), with each eddy acting independently and inducing a random walk of the tracer. The large scale eddy-diffusivity estimated from the FSLE plots was about $10^4$ $m^2/s$ in the northern and southern Bay, and approximately $6 \times 10^3$ $m^2/s$ in the Andaman Sea region. Interestingly, before the emergence of an eddy-diffusive regime, the finite size diffusion coefficient showed a similar power-law behavior in all regions of the Bay (exponent of 1.73 with cluster size). These estimates can be used as a guideline for ocean models, being run at a given resolution, that hope to capture the 
stirring at the surface of the Bay in an accurate manner.
		
Finally, from a global perspective of mixing in the Bay, we moved to the analysis of a single fresh water mixing episode. Guided by satellite salinity 
data in 2015, we demonstrated how an eddy in the western Bay helps preserve the identity of post-monsoonal fresh water from the Ganga-Brahmaputra river mouth.
In particular, FSLE and FTLE maps clearly delineated kinematic boundaries that aid in mixing within eddies, but prevent the intermingling of fresh water 
within an eddy with the saltier external Bay over the timescale of a month. Thus, while eddies stir the salinity field on a large scale, they also help
maintain the freshness of water trapped within themselves.
		
\section{Acknowledgement}
The authors would like to thank AVISO and JPL for making MADT-H-UV and SMAP salinity data freely available. The authors would like to express their gratitude to Dr. Debasis Sengupta, Dr. Anirban Guha and Dr. Amit Tandon for helpful discussions and technical advice. We also thank the Divecha Centre for Climate Change, IISc for financial support.
		
\bibliography{reference}

\end{document}